\documentclass[twocolumn]{aastex7}
\usepackage{graphicx}
\usepackage[caption=false]{subfig}
\usepackage{verbatim}
\usepackage{import}
\usepackage{CMNDSManual}
\usepackage{CMNDSSTD}
\usepackage{CMNDSSamples}
\usepackage{CMNDSBLSig}
\usepackage{CMNDSOffsets}
\usepackage{CMNDSIFU}
\usepackage{CMNDSClean}
\usepackage{CMNDSBPT}
\usepackage{CMNDSMerg}
\usepackage{CMNDSLineWidths}
\usepackage{CMNDSDeltaV}
\usepackage{CMNDSLLineMstar}
\usepackage{CMNDSMstarHost}
\usepackage{CMNDSSurfDens}

\shorttitle{Offset Broad Lines in MaNGA}
\shortauthors{Barrows et al.}

\begin{document}

\accepted{for publication in ApJ}

\title{Recoiling Black Hole Candidates from Spatially Offset Broad Emission Lines in MaNGA}

\author[0000-0002-6212-7328]{R. Scott Barrows}
\affiliation{Department of Astrophysical and Planetary Sciences, University of Colorado Boulder, Boulder, CO 80309, USA}
\email{Robert.Barrows@Colorado.edu}

\author{Julia M. Comerford}
\affiliation{Department of Astrophysical and Planetary Sciences, University of Colorado Boulder, Boulder, CO 80309, USA}
\email{Julie.Comerford@colorado.edu}

\author[0000-0003-2667-7645]{James Negus}
\affiliation{Cooperative Institute for Research in Environmental Sciences, University of Colorado Boulder, Boulder, CO, 80309, USA}
\email{James.Negus@colorado.edu}

\author[0000-0002-2713-0628]{Francisco Muller-Sanchez}
\affiliation{Department of Physics and Materials Science, The University of Memphis, 3720 Alumni Avenue, Memphis, TN 38152, USA}
\email{F.Muller.Sanchez@memphis.edu}

\correspondingauthor{R. Scott Barrows}
\email{Robert.Barrows@Colorado.edu}

\begin{abstract}

From the Mapping Nearby Galaxies at Apache Point Observatory (MaNGA) survey, we identify \combOffSZ~off-nuclear broad (FWHM $>$\,1000\,\uV) \ha~and/or \hb~emission line sources that indicate spatially offset active galactic nuclei (AGN) candidates. In addition to massive black holes (MBHs) in on-going galaxy mergers, this selection can also find MBHs that have been ejected from the host galaxy nucleus due to MBH binary coalescence and asymmetric gravitational wave emission or the dynamical `slingshot' mechanism. Recoiling/slingshot MBHs are predicted to affect co-evolution between MBHs and their host galaxies, and they are observational tracers of past binary MBH mergers and gravitational wave emission. This is the first systematic search through an integral field spectroscopy survey for ejected MBHs to enable uniform constraints on their surface densities. We find that \combOptPerc\% (\combOptSZ/\combOffSZ) have optical image counterparts consistent with galaxy stellar cores from infalling MBHs before the close binary MBH stage. The remaining \combNoOptPerc\% (\combNoOptSZ/\combOffSZ) have large broad line luminosities relative to their stellar core mass upper limits ($\sim$\,\LLineMstarRatioDiffRnd~times larger than for central AGN), suggesting merger-driven MBH accretion enhancements or potentially ejected MBHs. The signatures of AGN-ionized narrow emission lines for recoil/slingshot candidates are weaker by \PercDiffblrofas\%, which is consistent with the ejected MBH scenario. The broad line projected velocity offsets range from $\sim$\,10\,$-$\,600\,\uV~and suggest motion within the host galaxy potentials. Finally, the implied recoiling MBH surface density upper limit is consistent with predictions that assume random spin orientations in MBH binaries.

\end{abstract}

\keywords{Active galactic nuclei (16) --- Emission line galaxies (459) --- Galaxy mergers (608) --- Gravitational waves (678) --- Spectroscopy (1558) --- Supermassive black holes (1663)}

\section{Introduction}
\label{sec:intro}

Galaxy mergers are a potential route for co-evolution between galaxies and massive black holes (MBHs; $M_{\bullet}$\,$>$\,$10^{5}$\,\Msun) through induced star formation, randomization of stellar orbits, and MBH growth \citep[e.g.,][]{DiMatteo:2005,Springel:2005a,Capelo:2015}. Furthermore, observations suggest that merger-driven accretion onto MBHs (visible as active galactic nuclei; AGN) may have an important role in the overall AGN population \citep[e.g.,][]{Satyapal:2014,Weston:2017,Goulding:2018,Gao:2020}.

Since galaxy mergers result in two or more MBHs that are bound to a common gravitational potential, if the stellar cores hosting the MBHs lose sufficient angular momentum, they will eventually migrate toward the center of mass of the merger remnant. Moreover, if two MBHs form a bound binary system, they may eventually merge through the emission of gravitational waves. The final moment of coalescence will result in net gravitational wave emission that imparts a force (or recoil) on the merged MBH \citep[e.g.,][]{Peres:1962,Bekenstein:1973} and, if strong enough, may displace it from the remnant galaxy nucleus (e.g., \citealp{Campanelli:2007b,Gualandris:2008,Lousto:2011}; for a review see \citealt{Centrella:2010}). Alternatively, if the timescale for MBH binary coalescence is longer than the galaxy merger timescale, then a third MBH may enter the system and cause the lightest MBH to be (`slingshot') ejected by dynamical instabilities \citep[e.g.,][]{Hoffman:2007,Bonetti:2016,Bonetti:2018}.

The recoil and slingshot phenomena will have their own unique effects on the co-evolution of galaxies and MBHs. In particular, the MBH masses that are observed to correlate with host galaxy bulge properties \citep[e.g.,][]{Gebhardt00,Ferrarese2000,Gultekin:2009} will be lower due to ejection of MBHs or interrupted accretion \citep[e.g.,][]{Volonteri:2007,Blecha:2011}. The distribution of recoil or slingshot velocities will also introduce scatter into those relations \citep[e.g.,][]{Blecha:2008,Sijacki:2011}. Moreover, recoiling or slingshot MBH detections will impose constraints on MBH binary formation, spin alignment, coalescence rates, and gravitational wave emission.

Practically, detection of recoiling or slingshot MBHs requires them to be actively accreting as observable AGN (see \citealt{Komossa:2012} for a review of electromagnetic signatures). This scenario suggests a combination of four observables: 1) the AGN will have an unobscured broad line region (BLR; the region within which gas is gravitationally bound to the MBH and accelerated to velocities $>$\,1000\,\uV~that produce Doppler broadened emission lines; \citealp[e.g.,][]{Sulentic:2000,Gaskell:2009}); 2) if the acceleration from the recoil or slingshot is strong enough, the BLR may appear offset in velocity-space from the host galaxy systemic \citep[e.g.,][]{Bonning:2007,Komossa:2008,Volonteri:2008b,Bogdanovic09a}; 3) the BLR may have a detectable spatial offset from the host galaxy nucleus \citep[e.g.,][]{Blecha:2019}; and 4) the BLR may not be spatially coincident with a detectable AGN-ionized narrow line region (NLR) if the interstellar medium (ISM) gas through which the ejected MBH travels is too diffuse or if the NLR is too small \citep{Blecha:2013b}. Additionally, a galaxy hosting a slingshot (rather than recoiling) MBH is more likely to also have a nuclear AGN since at least one MBH will remain in the nucleus.

Multiple systematic searches for candidate recoiling or slingshot MBHs have identified BLRs of spectroscopically classified quasi-stellar objects (QSOs) that are kinematically displaced relative to the narrow emission line systems of their host galaxies (e.g., \citealp{Komossa08a,BL09,Bogdanovic09b,Dotti09,Shields09,Decarli2010,Robinson:2010,Barrows:2011,Steinhardt:2012,Tsai:2013,Markakis:2015}; with new sources continually discovered). However, the physical interpretation of these velocity offsets is not clear, and long-term monitoring \citep[notably, e.g., the systematic campaign described in][]{Eracleous:2012,Runnoe:2015,Runnoe:2017} has revealed that most (if not all) of these kinematic BLR displacements can be explained by accretion disk hotspots or warped accretion disks \citep[e.g.,][]{Eracleous95,EH03}.

Therefore, confirmation of an off-nuclear (i.e., spatially offset) BLR is a necessary step toward finding strong recoiling or slingshot MBH candidates. Indeed, several such candidates have been identified serendipitously or from targeted follow-up spectroscopy \citep[e.g.,][]{Koss:2014,Kalfountzou:2017,Chiaberge:2018,Jadhav:2021}. However, these examples represent a small and heterogeneous collection that can not be used to understand ejected MBHs as a population.

Systematically identifying strong candidates in large numbers requires surveys with spatially-resolved spectroscopy. This requirement is currently best satisfied by the \mangatitle~(\ma) survey \citep{Bundy:2015}, which was part of the fourth phase of the Sloan Digital Sky Survey (SDSS-IV) and is the largest integral field spectroscopy (IFS) survey to-date. Broad emission lines have previously been systematically detected in the centers of \ma~galaxies and their companions \citep{Steffen:2022}. \citet{Negus:2024} recently expanded on this by using the full extent of the \ma~IFS bundles to create a catalog of galaxies with broad emission lines detected in any spatial element (spaxel) of the \ma~data cubes. This present work uses that catalog for the first systematic search through an IFS survey for offset BLRs to find infalling AGN in mergers and recoiling/slingshot MBH candidates.

This paper is structured as follows: in Section \ref{sec:proc} we outline our procedure for selecting spatially offset BLR candidates, in Section \ref{sec:optical} we describe the identification of optical image counterparts to the BLRs, in Section \ref{sec:NLR} we quantify the presence of AGN-ionized narrow emission line gas in the host galaxies, in Section \ref{sec:nature} we discuss the possible physical nature of the offset BLRs, in Section \ref{sec:results} we explore the implications for merger-driven AGN triggering and detection of ejected MBHs, and in Section \ref{sec:conc} we synthesize the results of our analysis and interpretation. Throughout we assume a flat cosmology defined by the nine-year Wilkinson Microwave Anisotropy Probe observations \citep{Hinshaw:2013}: $H_{0}$\,$=$\,69.3\,km\,Mpc$^{-1}$\,s$^{-1}$ and $\Omega_{M}$\,$=$\,0.287.

\begin{figure}[ht!]
\includegraphics[width=0.48\textwidth]{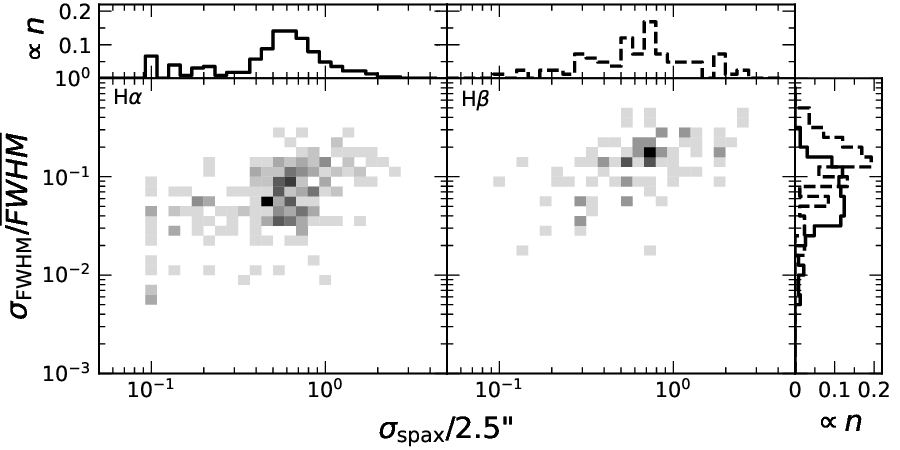}
\caption{\footnotesize{Linear-scale density map showing, for each galaxy, the standard deviations of the broad emission line spaxel FWHMs (normalized by the mean FWHM; \stdfwhm$/$\FWHMMean) against the standard deviations of the spaxel spatial positions (quadrature sum of the standard deviation of the spaxels in the X- and Y-dimensions) in units of the \ma~spatial resolution (\stdspax/2\farcs5) for \ha~(left) and \hb~(right) emission line detections. Only sources with more than one broad emission line spaxel are shown. The distributions of \stdfwhm$/$\FWHMMean~and \stdspax/2\farcs5 (normalized to a sum of unity) are shown along the right and top axes, respectively, for \ha~(solid) and \hb~(dashed). The concentration of abscissa values at \stdspax/2\farcs5\,$=$\,0.1 is due to sources with two adjacent broad emission line spaxels. Of the \stdfwhm$/$\FWHMMean~values, 100\% are $<$\,1 for both \ha~and \hb~detections. Of the \stdspax~values, \SIGSTDNormTwoPercha\%, \SIGSTDNormTwoPerchb\% (\ha, \hb~detections) are $<$\,2\,$\times$\,2\farcs5. These distributions show that the line widths and positions of the broad emission line detections in each galaxy are consistent and likely originate from a single source.}}
\label{fig:SPAXEL_FWHM_STD}
\end{figure}

\section{Procedure}
\label{sec:proc}

We begin with the catalog of \citet{Negus:2024} that consists of 1042 \ma~galaxies with detected broad spectral components in at least one spaxel. The broad component identifications were based on full-width at half-maximums of FWHM\,$>$\,1000\,\uV~associated with either of the H$\alpha$ or H$\beta$ emission lines (these are the broad emission lines spectrally accessible by \ma) and a detection significance of $>$\,5$\sigma$. The broad emission line components in \citet{Negus:2024} were fit with a single Gaussian component \citep[e.g.,][]{Oh:2015}. While AGN BLR emission may be modeled with more complex profiles \citep[e.g.,][]{Liu:2019} due to asymmetries from accretion disk hotspots or warped geometries, this choice does not alter the selection of candidate offset BLRs in this work. Therefore, we adopt these single component fits due to their uniformity for subsequent analysis and comparison.

\begin{figure}[ht!]
\includegraphics[width=0.48\textwidth]{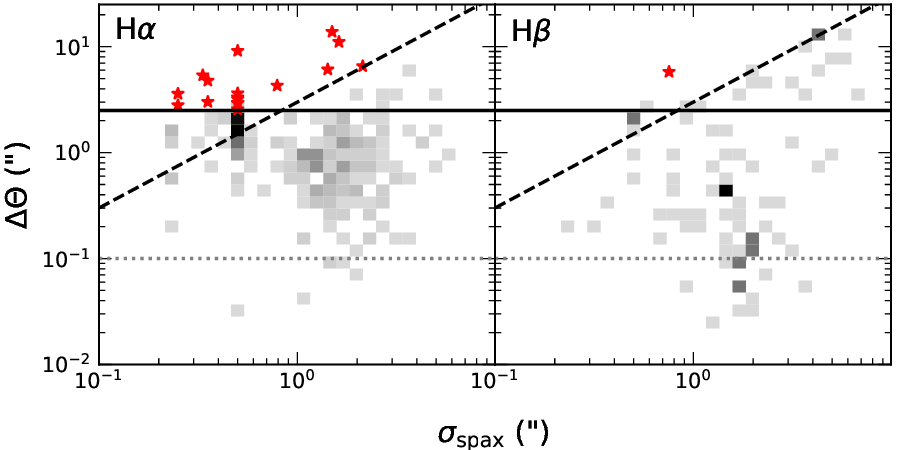}
\caption{\footnotesize{Angular offsets of \ha~(left) and \hb~(right) BLRs from the host galaxy centroids (\DeltaTheta) against the source extents (\stdspax; defined in Figure \ref{fig:SPAXEL_FWHM_STD}). The dotted gray line denotes the approximate \ma~absolute astrometric uncertainty of 0\farcs1. The dashed black lines denote the threshold of \DeltaTheta\,$>$\,\OffSigThresh\,$\times$\,\stdspax, and the solid black line denotes the additional uniform offset minimum of \DeltaTheta\,$>$\,2\farcs5 (Section \ref{sec:offset}). The parent sample is indicated by the linear-scale density map, while sources that pass the offset criteria are shown as red stars.  The concentrations at \stdspax\,$=$\,0\farcs25 and 0\farcs5 are due to sources with two adjacent spaxels and those with only one spaxel (for which \stdspax\,$:=$\,0\farcs5, i.e., the extent of one spaxel), respectively.}}
\label{fig:STD_DELTATHETA}
\end{figure}

In this work, we apply an additional threshold to the \citet{Negus:2024} catalog where the broad emission line to continuum flux density ratio is $>$\,5 times the standard deviation of the spectrum (over the fitting region), thereby accounting for the spectral quality. We have also removed cases where (assuming the host galaxy systemic redshift) multiple unidentified emission lines are observed, since this suggests the possibility that the broad emission line detections may originate from unrelated background or foreground sources. This yields \haBLSigSZ~and \hbBLSigSZ~galaxies with \ha~and \hb~broad emission line detections in at least one spaxel, respectively, among \combBLSigSZ~unique \ma~galaxies. We send this sample through the following steps to identify spatially offset BLR candidates.

\subsection{Spatial Centroids of Broad Emission Line Sources}
\label{sec:centroids}

For each galaxy, the standard deviations of the broad emission line spaxel positions (\stdspax) are shown in Figure \ref{fig:SPAXEL_FWHM_STD}. The values of \stdspax~are within two times the \ma~spatial resolution (2\farcs5) for \SIGSTDNormTwoPercha\%, \SIGSTDNormTwoPerchb\% (\ha, \hb~detections) of the sample. Therefore, at the \ma~resolution scale, no strong evidence that the broad emission line detections are from multiple sources within each galaxy is seen. Figure \ref{fig:SPAXEL_FWHM_STD} also shows the standard deviations of the spaxel broad emission line FWHMs (normalized by the mean FWHM: \stdfwhm$/$\FWHMMean). All values of \stdfwhm~are less than \FWHMMean~for \ha~and \hb~detections, thereby showing that the broad emission line widths are consistent within each galaxy.

The broad emission line detections are then assumed to originate from a single source and are referred to as BLRs. The BLR spatial centroids are taken to be the mean position of the broad emission line spaxels, weighted by the broad emission line flux. These values are computed separately for \ha~and \hb~detections.

\begin{figure}[ht!]
\includegraphics[width=0.48\textwidth]{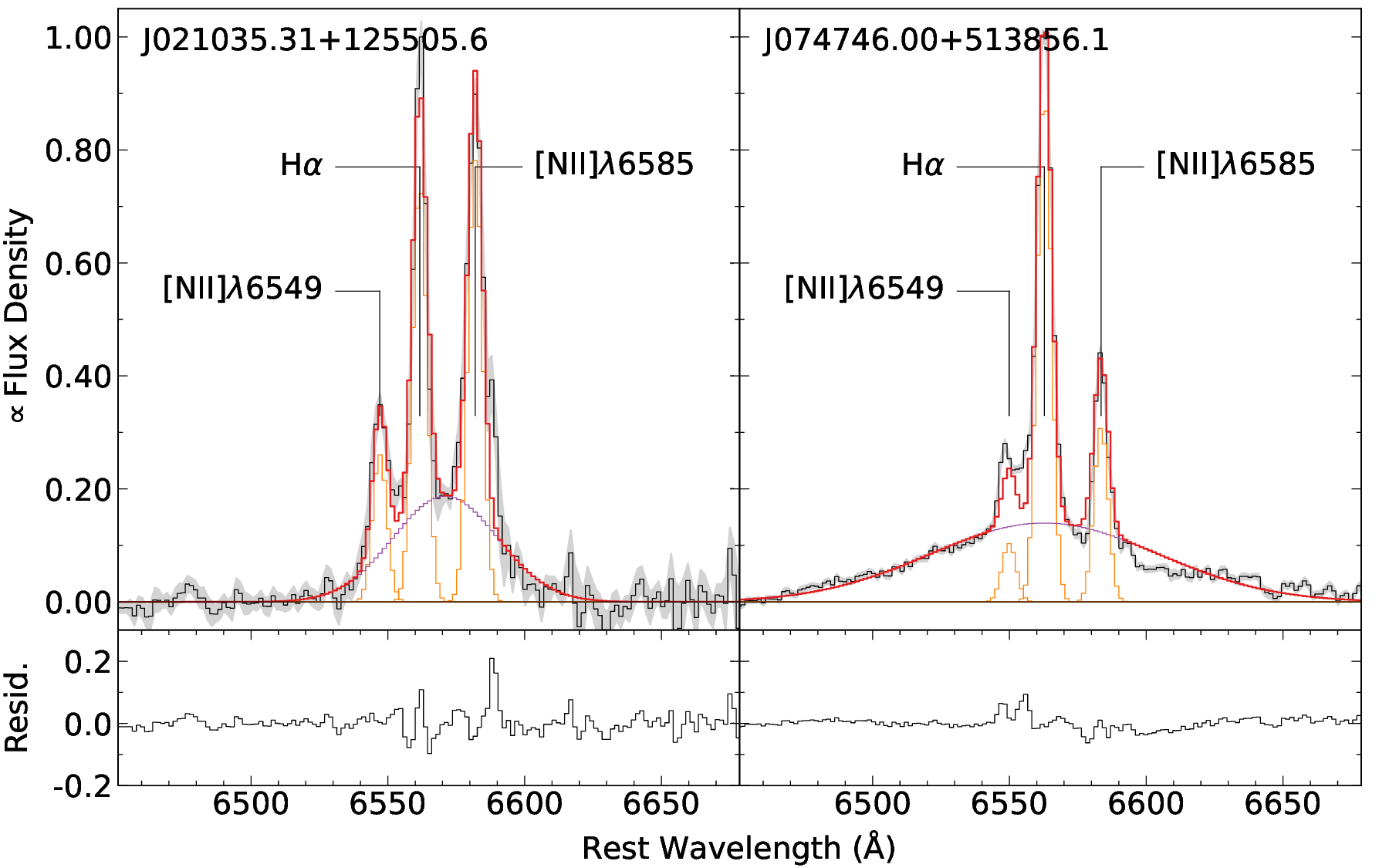}
\caption{\footnotesize{Best fits to the (continuum-subtracted) spectral regions (in the host galaxy rest-frame) where broad emission lines are detected. For clarity, each spectrum has been normalized by the peak flux. The narrow and broad line components are shown in orange and purple, respectively, and the model sum is shown in red. The uncertainty in the continuum subtracted spectrum is denoted by the gray-shaded region, and the residuals are shown at the bottom. Sorted by right ascension, these are the first two offset BLR sources in the sample. The remainder are shown and discussed in Section \ref{sec:spectral_models} of the Appendix (Figure \ref{fig:spectra_1}).}}
\label{fig:spectra_0}
\end{figure}

\subsection{Selection of Spatially Offset BLRs}
\label{sec:offset}

We select BLRs from Section \ref{sec:centroids} with angular offsets from the centroid of the \ma~target galaxy (\DeltaTheta; where the target galaxy centroid comes from the SDSS pipeline and is based on the $r$-band image) by $>$\,\OffSigThresh\,$\times$\,\stdspax. If a BLR has only one spaxel, then \stdspax\,$:=$\,1 spaxel. We also apply a uniform offset minimum threshold equal to the \ma~FWHM spatial resolution (\DeltaTheta\,$>$\,2\farcs5)\textbf{ so that the sources are not contaminated by beam smearing of nuclear emission}. The spatially offset selection is shown graphically in Figure \ref{fig:STD_DELTATHETA}. The \ma~absolute astrometry is tied to that of the SDSS ($\sim$\,0\farcs1; \citealp{Law:2016}) and is significantly smaller than the angular offsets required by this selection, so further image registration is not necessary. This yields a sample of \combOffSZ~unique spatially offset BLR candidates. One is detected in both lines, while the remaining 13 are detected only in \ha, which is consistent with \hb~being the weaker of the two recombination lines. Examples of the broad emission lines are shown in Figure \ref{fig:spectra_0}, and the remainder are shown in Section \ref{sec:spectral_models} of the Appendix. The full sample is listed in Table \ref{tab:Samp} and represents the largest compilation of spatially offset AGN that is spectroscopically selected from broad emission lines.

\begin{figure}[ht!]
\includegraphics[width=0.48\textwidth]{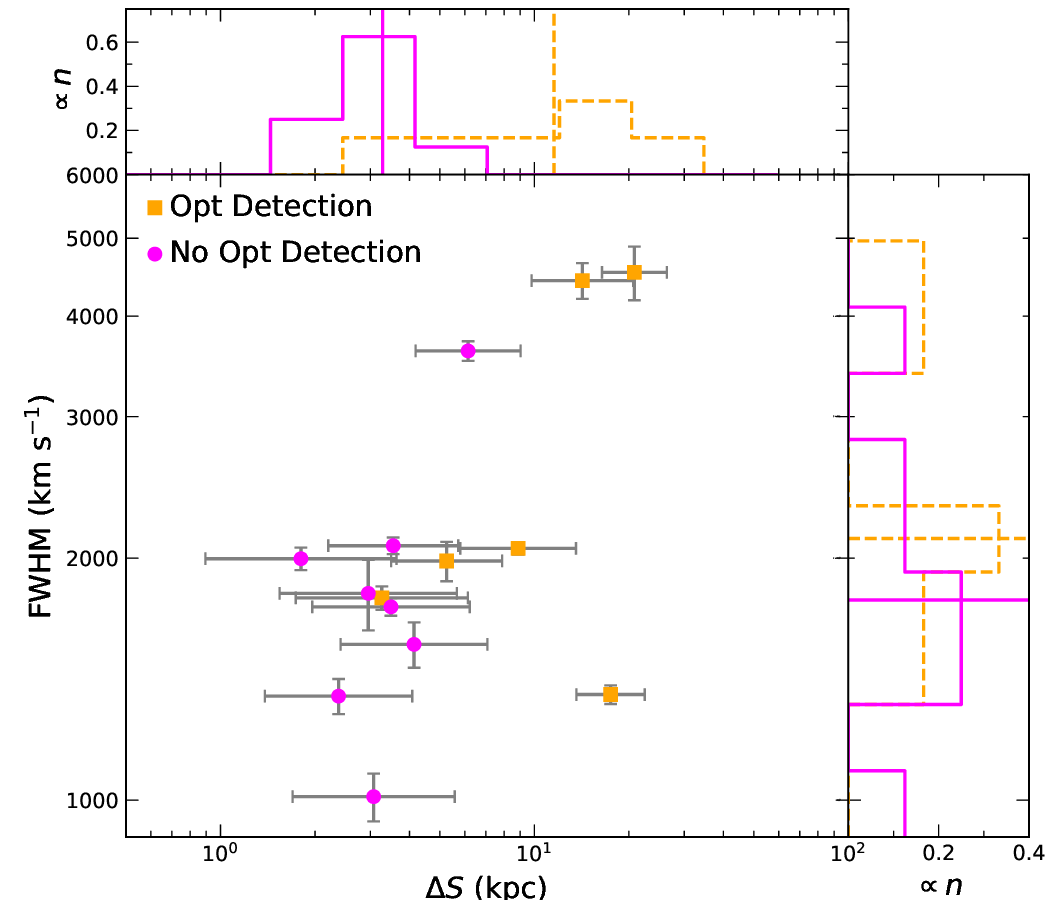}
\caption{\footnotesize{BLR full-width at half-maximum (FWHM; maximum value among all spaxels in each source) against projected physical offset from the host galaxy centroid (\DeltaS) for the final sample of spatially offset BLR candidates with (orange squares) and without (magenta circles) optical counterpart detections (Section \ref{sec:optical}). The distributions of FWHM and \DeltaS~(each normalized to a sum of unity) are shown along the right and top axes, respectively, for the subsets with (orange, dashed) and without (magenta, solid) optical counterparts, and their median values are indicated by straight lines. The distributions of FWHM and \DeltaS~are typical of AGN BLRs and late-stage galaxy mergers, respectively.}}
\label{fig:FWHM_DELTAS}
\end{figure}

The BLR FWHMs (maximum value among all broad emission line spaxels in each source) and projected physical offsets of the BLR positional centroids from the host galaxy nuclei (\DeltaS) are listed in Table \ref{tab:Samp}, and their bivariate distribution is shown in Figure \ref{fig:FWHM_DELTAS}. The FWHMs span a range that is typical of BLRs in AGN (FWHM\,$\approx$\,1000\,$-$\,5000\,\uV), and the values of \DeltaS~are characteristic of merging galaxy pairs (\DeltaS\,$\approx$\,2\,$-$\,20\,kpc). For comparison in our subsequent analyses, central BLRs are selected to have \DeltaTheta\,$<$\,\stdspax~and \DeltaTheta\,$<$\,2\farcs5.

\section{Optical Image Counterparts}
\label{sec:optical}

We use \gf~\citep{Peng:2010} to search for optical sources in the SDSS imaging that are spatially coincident with the BLR candidates from Section \ref{sec:proc}. Following \citet{Barrows:2016,Barrows:2018}, we first run \se~\citep{Bertin:Arnouts:1996} to obtain a list of significantly detected ($>$\,3\,$\sigma$) sources. We then use these source parameters to define input Sersic components for the \gf~models (Sersic profiles are appropriate for fitting to the stellar bulges expected to host AGN). We also run a model that additionally has a point spread function (PSF) component to test if its inclusion improves the model based on an $F$-test (in no cases do we find a PSF component warrants inclusion). 

We consider an SDSS image source to be associated with an offset BLR candidate if their respective spatial centroids are in agreement within 2\farcs5\,$+$\,\stdspax~(to be consistent with the selection of spatially offset AGN in Section \ref{sec:offset}). The stellar mass for each optical counterpart is estimated by applying the mass ratio between it and the host galaxy (from the \gf~modeling) to the host galaxy stellar mass (from \texttt{Pipe3D}; \citealp{Lacerda:2022,Sanchez:2022}). For non-detections, upper limits correspond to 5 times the local background flux. Figures \ref{fig:panels_comb_opt} and \ref{fig:panels_comb_no_opt_recoil} show galaxy images and \ma~maps for the full samples of spatially offset H$\alpha$ and H$\beta$ BLR candidates with (\combOptSZ) and without (\combNoOptSZ) optical image counterpart detections, respectively. The counterpart stellar masses (or upper limits) are listed in Table \ref{tab:Samp}.

The distribution of \DeltaS~values for the subset with optical image counterparts is biased toward larger values compared to the subset without. This difference is likely due to the larger angular sizes of the \ma~IFS hexagon bundles for the subset with image detections (mean diameter of \IFUSzASMeano$''$ compared to \IFUSzASMeanr$''$; see Figures \ref{fig:panels_comb_opt} and \ref{fig:panels_comb_no_opt_recoil}).

\begin{deluxetable*}{cccccccccccccc}
\tabletypesize{\footnotesize}
\tablecolumns{14}
\tablecaption{Spatially Offset Broad Emission Line Sources}
\tablehead{
\colhead{Host Galaxy}  &
\colhead{$z$} &
\colhead{W} &
\colhead{$L_{BL}$} &
\colhead{W$_{\rm{[OIII]}}$} &
\colhead{Sig$_{\rm{[OIII]}}$} &
\colhead{\DeltaS} &
\colhead{\DeltaV} &
\colhead{\MstarHost} &
\colhead{\MstarBL} &
\colhead{\FABL} &
\colhead{\FAN} &
\colhead{BL} &
\colhead{Det} \\
\colhead{($-$)} &
\colhead{($-$)} &
\colhead{(\uVFrac)} &
\colhead{(log[\uLumFrac])} &
\colhead{(\uVFrac)} &
\colhead{($-$)} &
\colhead{(kpc)} &
\colhead{(\uVFrac)} &
\colhead{(log[\Msun])} &
\colhead{(log[\Msun])} &
\colhead{$-$} &
\colhead{$-$} &
\colhead{($-$)} &
\colhead{($-$)} \\
\colhead{1} &
\colhead{2} &
\colhead{3} &
\colhead{4} &
\colhead{5} &
\colhead{6} &
\colhead{7} &
\colhead{8} &
\colhead{9} &
\colhead{10} &
\colhead{11} &
\colhead{12} &
\colhead{13} &
\colhead{14}
}
\startdata
J021035.31+125505.6 & 0.0996 & $2056\pm{35}$ & $39.83^{+0.04}_{-0.04}$ & 373 & 0.1 & $8.9^{+4.7}_{-3.1}$ & $-292^{+596}_{-196}$ & $11.42\pm0.09$ & $11.1\pm0.7$ & 1.00,0.50 & 0.87,0.57 & H$\alpha$ & 1 \\ 
J074746.00+513856.1 & 0.1009 & $4534\pm{347}$ & $42.02^{+0.14}_{-0.21}$ & 292 & 2.5 & $20.8^{+5.6}_{-4.4}$ & $-4^{+103}_{-4}$ & $11.54\pm0.09$ & $10.8\pm0.7$ & 0.70,0.72 & 0.94,0.88 & H$\alpha$ & 1 \\ 
J090401.02+012729.1 & 0.0534 & $1739\pm{43}$ & $38.78^{+0.05}_{-0.06}$ & 256 & 1.0 & $3.5^{+2.7}_{-1.5}$ & $45^{+105}_{-31}$ & $11.02\pm0.09$ & 9.7$^a$ & 0.87,0.67 & 0.21,0.14 & H$\alpha$ & 0 \\ 
J094630.90+345500.6 & 0.0414 & $2072\pm{49}$ & $38.84^{+0.05}_{-0.05}$ & 267 & 1.0 & $3.5^{+2.2}_{-1.3}$ & $-14^{+103}_{-12}$ & $10.99\pm0.06$ & 9.2$^a$ & 0.00,0.00 & 0.00,0.00 & H$\alpha$ & 0 \\ 
J103723.62+021845.5 & 0.0402 & $1984\pm{112}$ & $39.31^{+0.11}_{-0.14}$ & 318 & 6.8 & $5.3^{+2.6}_{-1.8}$ & $58^{+105}_{-37}$ & $11.22\pm0.08$ & $10.8\pm0.8$ & 0.45,0.50 & 0.78,0.98 & H$\alpha$ & 1 \\ 
J112458.71+470836.6 & 0.0537 & $1808\pm{181}$ & $38.87^{+0.18}_{-0.30}$ & 299 & 0.2 & $3.0^{+2.7}_{-1.4}$ & $-28^{+106}_{-22}$ & $11.26\pm0.08$ & 9.5$^a$ & 0.00,0.00 & 0.00,0.06 & H$\alpha$ & 0 \\ 
J133938.88+272416.5 & 0.0351 & $1996\pm{64}$ & $38.08^{+0.06}_{-0.08}$ & 349 & 0.2 & $1.8^{+1.8}_{-0.9}$ & $611^{+1210}_{-406}$ & $10.97\pm0.07$ & 9.1$^a$ & 0.00,0.07 & 0.42,0.35 & H$\alpha$ & 0 \\ 
J135638.55+433508.7 & 0.1031 & $1353\pm{36}$ & $39.35^{+0.05}_{-0.06}$ & 258 & 0.1 & $17.5^{+5.0}_{-3.9}$ & $-39^{+103}_{-28}$ & $11.36\pm0.09$ & $10.9\pm0.7$ & 0.00,0.00 & 0.00,0.01 & H$\alpha$ & 1 \\ 
J150847.79+301014.0 & 0.0589 & $1562\pm{101}$ & $38.92^{+0.12}_{-0.17}$ & 261 & 0.1 & $4.1^{+2.9}_{-1.7}$ & $-31^{+103}_{-24}$ & $11.54\pm0.08$ & 9.7$^a$ & 0.00,0.00 & 0.00,0.00 & H$\alpha$ & 0 \\ 
J160400.93+255719.4 & 0.0486 & $1010\pm{68}$ & $38.26^{+0.13}_{-0.18}$ & 239 & 0.0 & $3.1^{+2.5}_{-1.4}$ & $-32^{+107}_{-25}$ & $11.32\pm0.06$ & 9.4$^a$ & 0.00,0.00 & 0.00,0.00 & H$\alpha$ & 0 \\ 
J161114.40+241330.0 & 0.0325 & $1347\pm{67}$ & $37.98^{+0.10}_{-0.13}$ & 230 & 0.1 & $2.4^{+1.7}_{-1.0}$ & $-91^{+105}_{-49}$ & $10.82\pm0.06$ & 9.0$^a$ & 0.00,0.00 & 0.00,0.00 & H$\alpha$ & 0 \\ 
J163233.80+262250.6 & 0.0586 & $3620\pm{101}$ & $39.77^{+0.06}_{-0.07}$ & 234 & 0.6 & $6.1^{+2.9}_{-2.0}$ & $50^{+103}_{-34}$ & $11.51\pm0.07$ & 9.7$^a$ & 0.00,0.00 & 0.44,0.32 & H$\alpha$ & 0 \\ 
J170450.70+344858.9 & 0.0564 & $1784\pm{59}$ & $38.72^{+0.07}_{-0.08}$ & 253 & 0.9 & $3.3^{+2.9}_{-1.5}$ & $-11^{+102}_{-10}$ & $11.39\pm0.07$ & $11.2\pm0.8$ & 0.00,0.00 & 0.00,0.00 & H$\alpha$ & 1 \\ 
J211326.02+005828.0 & 0.1374 & $4429\pm{226}$ & $40.48^{+0.10}_{-0.13}$ & 1310 & 0.6 & $14.2^{+6.4}_{-4.4}$ & $368^{+1275}_{-285}$ & $11.66\pm0.08$ & $10.9\pm0.8$ & 0.00,0.00 & 0.21,0.04 & H$\alpha$, H$\beta$ & 1
\enddata
\tablecomments{Column 1: Host galaxy SDSS name; column 2: host galaxy redshift; column 3: maximum full-width at half-maximum (FWHM) of the broad emission line spaxel detections for the source; column 4: integrated broad emission line luminosity from all of the broad emission line spaxels; columns 5-6: maximum FWHM and detection significance of the outflow component (among all spaxels associated with a BLR candidate), as measured from the \oiiilambda~emission line; column 7: projected physical offset of the BLR positional centroid from the host galaxy centroid; column 8: projected velocity offset of the BLR (flux-weighted mean of the broad emission line spaxels) from the host galaxy systemic velocity (defined as \DeltaV\,$=c\times(\lambda_{\rm{vac}}-\lambda_{BL})/\lambda_{\rm{vac}}$); column 9: stellar mass of the host galaxy; column 10: stellar mass or upper limit of the BLR optical image counterpart; columns 11-12: AGN spaxel fractions associated with the broad emission line source and the host galaxy nucleus based on the \nii,\sii~BPT diagrams; column 13: recombination line in which the BLR detection is made; and column 14: whether an SDSS image counterpart is detected (1) or not (0).\\ $^a$Upper limit.}
\label{tab:Samp}
\end{deluxetable*}

\section{Narrow Emission Line Diagnostics}
\label{sec:NLR}

To quantify the presence of extended host galaxy NLR gas that has been photo-ionized by the AGN, we examine the locations of the \ma~spaxels on the `Baldwin-Phillips-Terlevich' (or `BPT') diagram. We use the calibration from \citet{Kewley:2006} based on the line ratios of \oiiilambda/\hb~versus \niilambdaTwo/\ha~(\nii~BPT diagram) and versus \siilambdaOneTwo/\ha~(\sii~BPT diagram), where the narrow line fluxes are taken from the \ma~Data Analysis Pipeline and are required to have signal-to-noise ratio $>$\,3.

Figures \ref{fig:panels_comb_opt} and \ref{fig:panels_comb_no_opt_recoil} show the locations of the spaxels within a box region of size 2\,$\times$\,\stdspax~around the BLR centroid on the \nii~and \sii~BPT diagrams. Since the \ma~pipeline spectral fitting does not include broad components, we omit from these regions the spaxels with broad emission line detections (from \citealt{Negus:2024}) which might affect the narrow emission line measurements. However, in our subsequent analyses, we test the effects of including these spaxels and find that the results are qualitatively unchanged (Section \ref{sec:BPT}).

Following \citet{Wylezalek:2018}, we determine the fraction of these spaxels (within the box region described above) that contain `AGN' emission based on the \nii~BPT diagram (\FANBL) and `Seyfert' or `LINER' (low-ionization nuclear emission line region) emission based on the \sii~BPT diagram (\FASBL). We also compute the AGN spaxel fraction centered on the nucleus of each host galaxy (\FANN~and \FASN), using the same box region as described above for uniform comparison. Values of \FABL~and \FAN~are listed in Table \ref{tab:Samp}.

\begin{figure*}[ht!]
\includegraphics[width=0.96\textwidth]{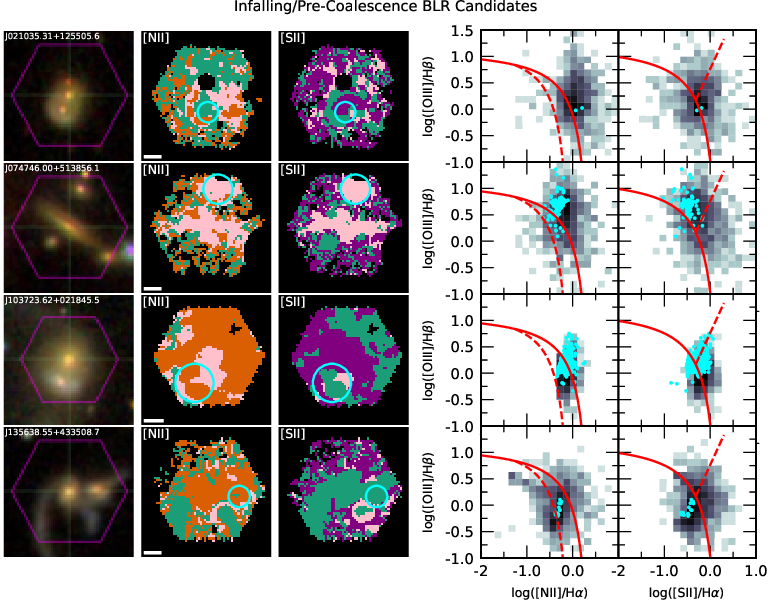}
\caption{\footnotesize{Full sample of \ma~galaxies with spatially offset BLRs coincident with optical detections from the SDSS imaging. For each galaxy, the panels contain (from left to right) the SDSS $g$\,$+$\,$r$\,$+$\,$i$ color composite images (with the \ma~IFS fiber bundle hexagon shown in magenta), the \ma~IFS maps and BPT classifications using the \nii-based criteria (pink\,$=$\,`AGN', orange\,$=$\,`composite', and green\,$=$\,`star forming') and \sii-based criteria (pink\,$=$\,`Seyfert', purple\,$=$\,`LINER', and green\,$=$\,`star forming'), and the \nii~and \sii~BPT diagrams. In the BPT plots, the solid red lines denote the pure AGN demarcation, while the dashed red lines denote the pure star formation (\nii) and LINER (\sii) demarcations. The log-scale density map denotes spaxels from the full \ma~map (only for spaxels with BPT classifications). The cyan dots denote the subset of the spaxels associated with the BLRs, selected to be within a box (centered on the BLR centroid) of size 2\,$\times$\,\stdspax~(to include extended narrow line region emission around the BLR) but not including the spaxels that have detected broad emission lines (since the \ma~pipeline does not account for them). The BLR positions are indicated by the cyan circles with radii of 2\farcs5\,$+$\,\stdspax~(radius used for matching with optical counterparts; Section \ref{sec:optical}). The white scale bar indicates 4$''$.}}
\label{fig:panels_comb_opt}
\end{figure*}

\begin{figure*}[ht!]
\ContinuedFloat
\includegraphics[width=0.96\textwidth]{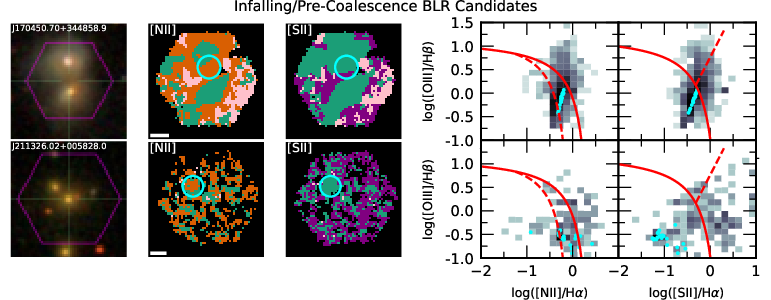}
\caption{\footnotesize{Continued}}
\end{figure*}

\begin{figure*}[ht!]
\includegraphics[width=0.96\textwidth]{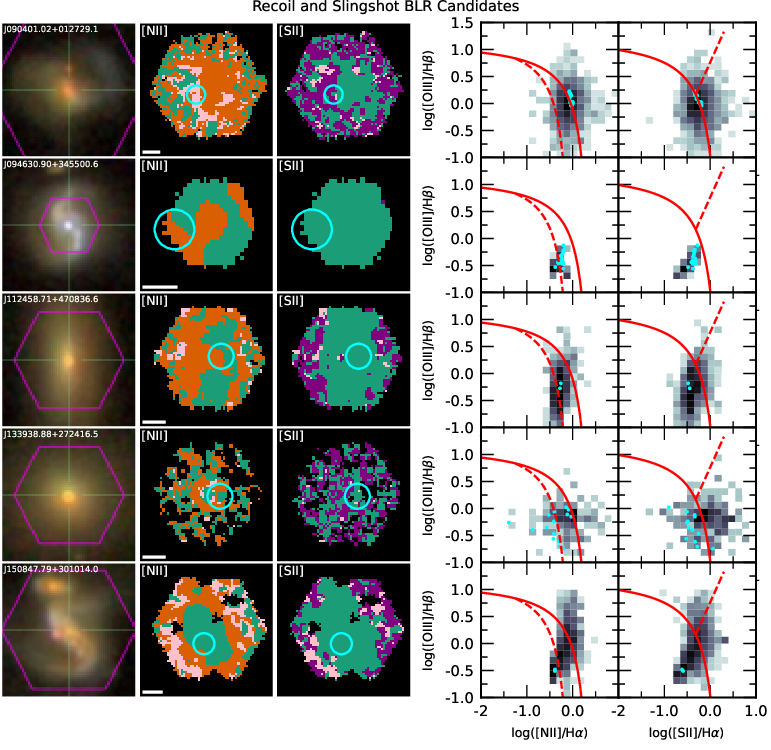}
\caption{\footnotesize{Same as Figure \ref{fig:panels_comb_opt} but for the full sample without optical image counterpart detections.}}
\label{fig:panels_comb_no_opt_recoil}
\end{figure*}

\begin{figure*}[ht!]
\ContinuedFloat
\includegraphics[width=0.96\textwidth]{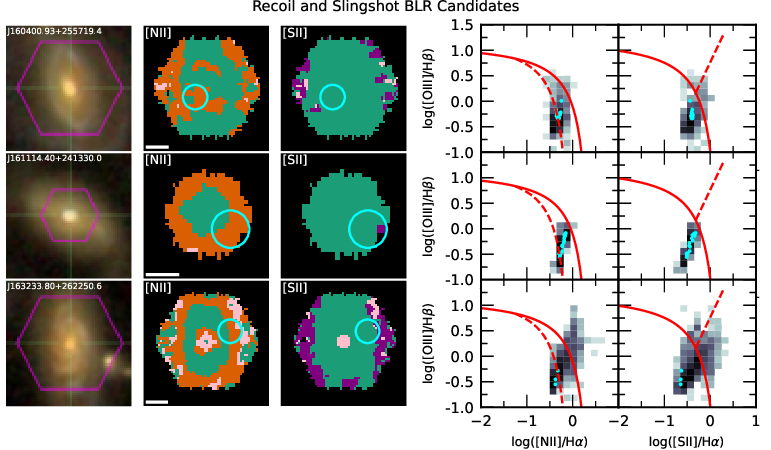}
\caption{\footnotesize{Continued.}}
\end{figure*}

\section{Nature of the Spatially Offset BLRs}
\label{sec:nature}

In the following sections we discuss the possible nature of the spatially offset BLRs selected in this work. The discussions are presented within the context of unrelated blackground or foreground sources (Section \ref{sec:cont}), outflows (Section \ref{sec:outflows}), supernovae (Section \ref{sec:SN}), infalling MBHs in pre-coalescence mergers (Section \ref{sec:infalling}), and recoiling or slingshot MBHs (Section \ref{sec:recoil}).

\subsection{Background or Foreground Contaminants}
\label{sec:cont}

Broad emission line detections in the spaxels associated with \ma~galaxies may be emitted from background or foreground AGN and imprinted on the spectra of the galaxies. Such unrelated broad emission lines can mimic lines in the target galaxy if their observed wavelengths are near those of \ha~or \hb~in the target rest frame. Galaxies with multiple unidentified emission lines are already removed from the sample (Section \ref{sec:proc}). However, in cases where only one broad emission line is detected near the wavelengths of \ha~or \hb~in the target rest frame, the possibility of it being emitted by an unrelated source can be neither confirmed nor rejected.

To statistically quantify this possibility, we compute the number of broad emission line AGN (based on the 16th SDSS Data Release QSO catalog; \citealp{Lyke:2020}) that would randomly be projected onto the \ma~map for each offset AGN in our sample. The QSO redshift ranges we use are from $z$\,$=$\,0 to the redshift at which observed \civ~emission from a background QSO equals that of \ha~in the offset BLR candidate host galaxy. Assuming any unrelated broad emission lines would be from QSOs,  the highest probability of one randomly appearing within a \ma~map from our sample is $2.7\times10^{-4}$.

\subsection{Outflows}
\label{sec:outflows}

While the broad line selection adopted for this work echoes numerous studies finding that the threshold of 1000\,\uV~cleanly identifies broad line regions in \ha~and \hb~emission around AGN \citep[e.g.,][]{Hao:2005,Schneider:2010,Shen:2011a,Stern:Laor:2012,Oh:2015,Liu:2019}, blending of multiple emission lines (i.e., in the \ha/\nii~complex) that are broadened by outflows (driven by central AGN or star formation; e.g., \citealp{Wylezalek:2020,Meena:2021}) can potentially mimic the presence of a single broad component in the line fitting procedure. To quantify the possibility of outflows in this work, independently of the \ha~complex, we test models of the continuum-subtracted \oiiilambda~emission line in which an additional Gaussian component is added, with an unconstrained line width intended to account for potential AGN outflows that are often seen in lines of high-ionization potential in AGN \citep[e.g.,][]{Komossa:2008b}. We fit these models on the spectra of all spaxels in which broad \ha~components are detected.

For each BLR candidate, the maximum FWHM value and maximum detection significance of the potential outflow components are listed in Table \ref{tab:Samp}. In only one source is an additional Gaussian present at a significant ($>$\,3$\sigma$) level (J1037), and the FWHM is 318\,\uV~(compared to 1984\,\uV~for that of the broad \ha~component). Hence, no evidence is seen for outflows that are broad enough to mimic the detected broad emission line components. This result is consistent with 13/\combOffSZ~of the broad emission line FWHMs in our sample exceeding the maximum outflow values observed in \ma~galaxies ($\sim$1200\,\uV; based on \oiiilambdaall~emission line widths; \citealp{Wylezalek:2020}), as shown in Figure \ref{fig:FWHM_DELTAS}.

\subsection{Supernovae}
\label{sec:SN}

Hydrogren-rich supernovae (SNe) are occasionally observed to have broad Balmer emission lines with FWHMs up to several 1000\,\uV \citep[e.g.,][]{Yan:2015,Kokubo:2019,Kuncarayakti:2023}. The highly transient nature of SNe suggests that our procedure is unlikely to preferentially select them, though this possibility can not be ruled out for any individual sources. For the offset AGN candidates with optical image counterpart detections, the counterpart stellar masses strongly suggest the BLRs are associated with MBHs in galaxy nuclei, hence disfavoring the SNe explanation in these cases. However, for those without image counterpart detections (\combNoOptPerc\%), the SNe possibility can not be definitively rejected with the current data. Multi-epoch spectra is the most effective means of constraining this possibility by determining whether or not the broad line emission is persistent.

\subsection{\text{Offset AGN}}
\label{sec:infalling}

The masses of the hypercompact stellar systems (HCSSs; i.e., tightly gravitationally bound stellar cores) that are predicted to accompany recoiling MBHs \citep[e.g.,][]{Gualandris:2008,Komossa:2008c,Merritt:2009,Li:2012,Lena:2020} are proportional to the MBH masses and velocity dispersions of the galaxy stellar cores where the recoils occur, and they are inversely proportional to the recoil kick velocities. Even assuming a relatively low kick velocity of 100\,\uV~\citep[e.g.,][]{Schnittman:2007,Blecha:2008} and a typical Milky Way bulge stellar velocity dispersion of 100\,\uV~\citep[e.g.,][]{Valenti:2018}, the stellar masses of the \combOptSZ~optical image counterpart detections in this work (6.3\,$\times$\,$10^{10}$\,$-$\,1.6\,$\times$\,$10^{11}$\,\Msun) are all significantly larger than their predicted HCSS masses ($\sim$\,$10^5$\,$-$\,$10^7$\,\Msun; using the derivation from \citealt{Merritt:2009}, assuming collisional loss cone repopulation, and using the MBH mass estimates from \citealt{Negus:2024}). The stellar cores around slingshot MBHs will be even smaller since MBHs will remain in the host galaxy nucleus. Therefore, the offset BLR candidates with optical image counterpart detections (Figure \ref{fig:panels_comb_opt}) are unlikely to be from ejected MBHs, and they are instead likely infalling (i.e., pre-coalescence) AGN in galaxy mergers or merger remnants \citep[e.g.,][]{Mueller-Sanchez:2015,Barrows:2017,Skipper:2018,Stemo:2021,Barrows:2024}.

However, the \combNoOptSZ~offset BLRs without optical image counterpart detections (Figure \ref{fig:panels_comb_no_opt_recoil}) remain recoil or slingshot candidates since, due to the absence of lower limits, their hypothetical stellar core masses are potentially consistent with values predicted for HCSSs. Indeed, using the SDSS galaxy morphological classifications of \citet{Nevin:2023} and a merger probability threshold of 0.5 \citep[i.e.,][]{Comerford:2024}, \PostCoalPerc\% of them are classified as post-coalescence mergers (as expected for the ejected MBH scenario).

\subsection{Recoiling and Slingshot MBHs}
\label{sec:recoil}

The recoil or slingshot scenarios can be neither confirmed nor rejected with the current data for the \combNoOptSZ~offset BLR candidates without optical image counterparts. Therefore, to further constrain the likely nature of these sources, in this section we compare the optical extended AGN-ionized emission (Section \ref{sec:BPT}) and the velocity offsets (Section \ref{sec:DeltaV}) to expectations for recoiling and slingshot MBHs.

\subsubsection{Spatially Extended AGN Narrow Line Emission}
\label{sec:BPT}

\begin{figure*}[ht!]
$
\begin{array}{c c}
\includegraphics[width=0.48\textwidth]{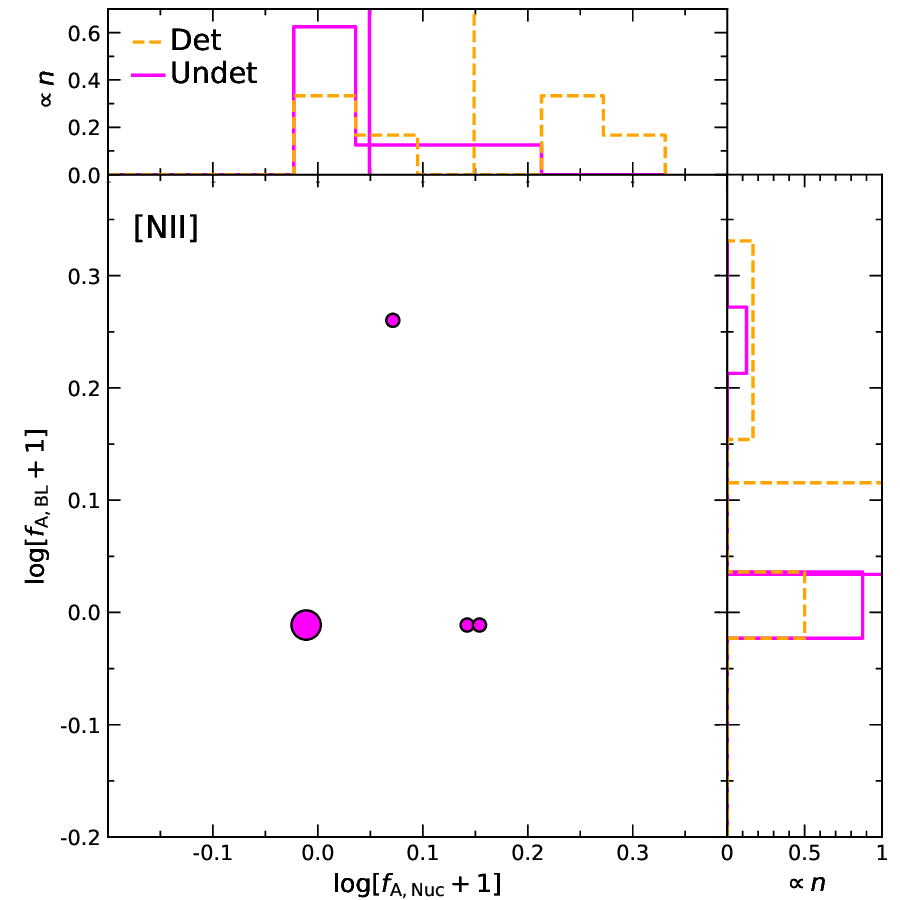} &
\includegraphics[width=0.48\textwidth]{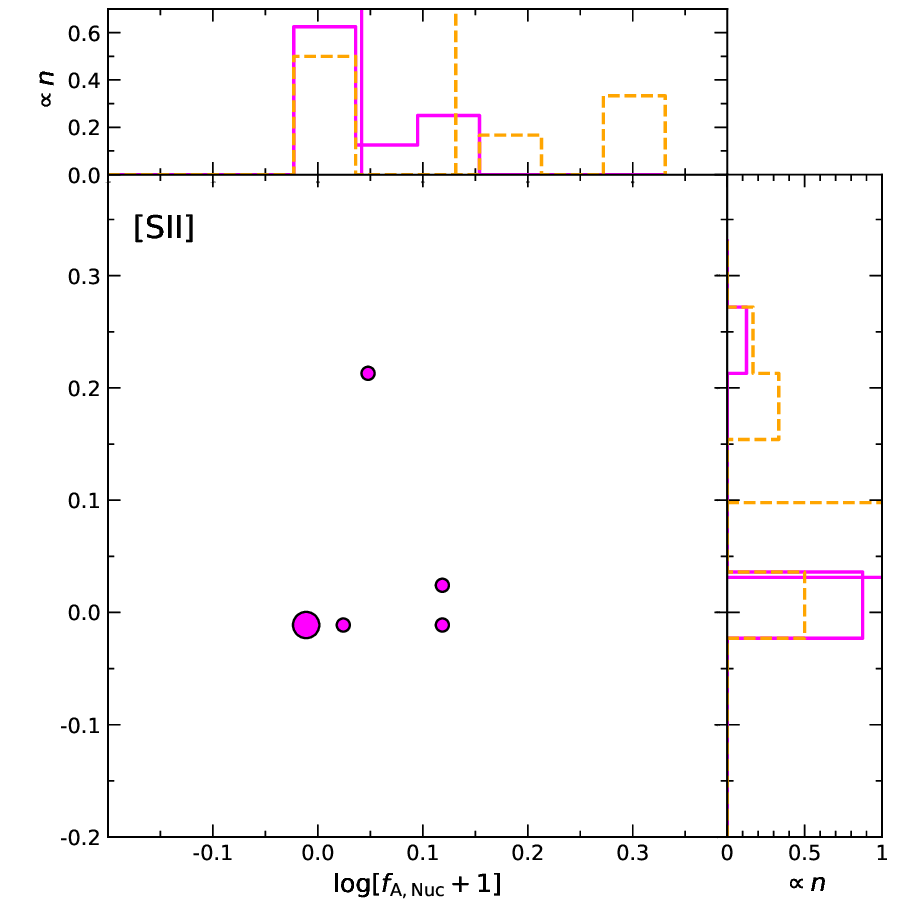}
\end{array}
$
\caption{\footnotesize{AGN BPT spaxel fractions of the BLRs (\FABL) against those of the host galaxy nuclei (\FAN) for AGN selections based on the \nii~(left) and \sii~(right) BPT diagrams for the offset AGN without optical image counterpart detections (i.e., the ejected AGN candidates; magenta circles). See Section \ref{sec:NLR} for details on the AGN spaxel fraction calculations. Since sources cluster at \FABL\,$=$\,\FAN\,$=$\,0, marker sizes indicate the number of sources: small and large circles indicate 1 and 5 sources, respectively (left panel), and 1 and 4 sources, respectively (right panel). The distributions of \FABL~and \FAN~are shown along the right and top axes, respectively, normalized to a sum of unity. For comparison, the distributions for the offset AGN with optical image counterparts (i.e., likely infalling AGN; orange dashed) are also shown. The mean values are indicated with straight lines of corresponding color and style. The mean values of \FABL~are smallest for the ejected AGN candidates (consistent with the recoil/slingshot scenario). The mean values of \FAN~are also smaller for the ejected AGN candidates, which is more consistent with the recoil (as opposed to slingshot) scenario.}}
\label{fig:F_AGN_bl_nuc}
\end{figure*}

If the ISM gas in the local environment of an ejected AGN is too diffuse, or if the AGN is too faint, an AGN-ionized NLR at the offset BLR position will not be detected \citep{Blecha:2013b}. In this scenario, sources with \FABL~values close to unity may be particularly weak recoil or slingshot candidates, while those with \FABL\,$=$\,0 are the stronger candidates. Indeed, the \FABL~values for the offset BLRs without optical image counterparts are \PercDiffblrofan\% and \PercDiffblrofas\% lower (for the \nii~and \sii~diagnostics, respectively) than those with counterparts (Figure \ref{fig:F_AGN_bl_nuc}). Moreover, the Kolmogorov-Smirnov (KS) statistic and null hypothesis probability are \KSStatblrofan~and \KSPblrofan, respectively (for both the \nii~and \sii~diagnostics), that they have similar \FABL~distributions. This result is qualitatively unchanged when including the spaxels with broad emission lines (see Section \ref{sec:NLR}), and may provide some evidence in support of the recoil or slingshot scenarios. However, relatively low values of \FABL~among offset BLRs without counterparts are potentially explained by their lower AGN luminosities (the median luminosity is \LLineRatioDiffInt~times smaller than for those with optical counterparts; Table \ref{tab:Samp}).

AGN emission in the host galaxy nucleus may occur after a MBH slingshot if one or both of the remaining nuclear MBHs (or a coalesced MBH remnant) is actively accreting. Extended photo-ionized AGN emission may also be present in the nucleus where a recoil occurred if an AGN was previously there and its ionization signatures remain \citep[e.g.,][]{Keel:2015,LaMassa:2015,Runnoe:2016,Comerford:2017b}. However, such emission is likely to have faded because the time that recoiling AGN are predicted to spend at the observed values of \DeltaS~for our sample range from $\sim$1\,$-$\,10\,Myrs \citep{Blecha:2016}, while light echo timescales for AGN-ionized optical emission are estimated to be only $\sim$\,$10^4$\,$-$\,$10^5$\,yrs for extended NLRs \citep[e.g.,][]{Schawinski:2015}.  Therefore, the ejected AGN candidates with \FABL\,$=$\,0 and \FAN\,$>$0 may be stronger slingshot candidates, while those with \FABL\,$=$\,\FAN\,$=$\,0 may be stronger recoil candidates.

Obscuration would also attenuate NLR AGN emission in the host galaxy nuclei. However, when converting the Balmer decrements of the host galaxy nuclei into color excesses (\ebv; using the relation from \citealt{Yuan:2018}), the host galaxy nuclei with \FAN\,$=$\,0 have a mean \ebv~value of only 0.41, compared to the mean of \BalmerDecMeanr~for the full sample. So there is no systematic evidence for a lack of nuclear AGN NLR emission being caused by obscuration.

\subsubsection{Velocity Offsets}
\label{sec:DeltaV}

Theoretical work predicts that recoil (`kick') velocities strongly depend on the relative spin alignment of the progenitor MBHs, with relatively small kick velocities generated for more aligned MBH spins  \citep[e.g.,][]{Campanelli:2007a,Campanelli:2007b,Schnittman:2007}. The distribution of kick velocities will determine the observed distribution of recoiling MBH velocities, modulo the effects of damping due to dynamical friction \citep[e.g.,][]{Blecha:2011,Sijacki:2011}, oscillation within the host galaxy (assuming the kick velocity is below the galactic escape velocity; e.g., \citealp{Choksi:2017}), and projection onto the sky plane.

\begin{figure}[ht!]
\includegraphics[width=0.48\textwidth]{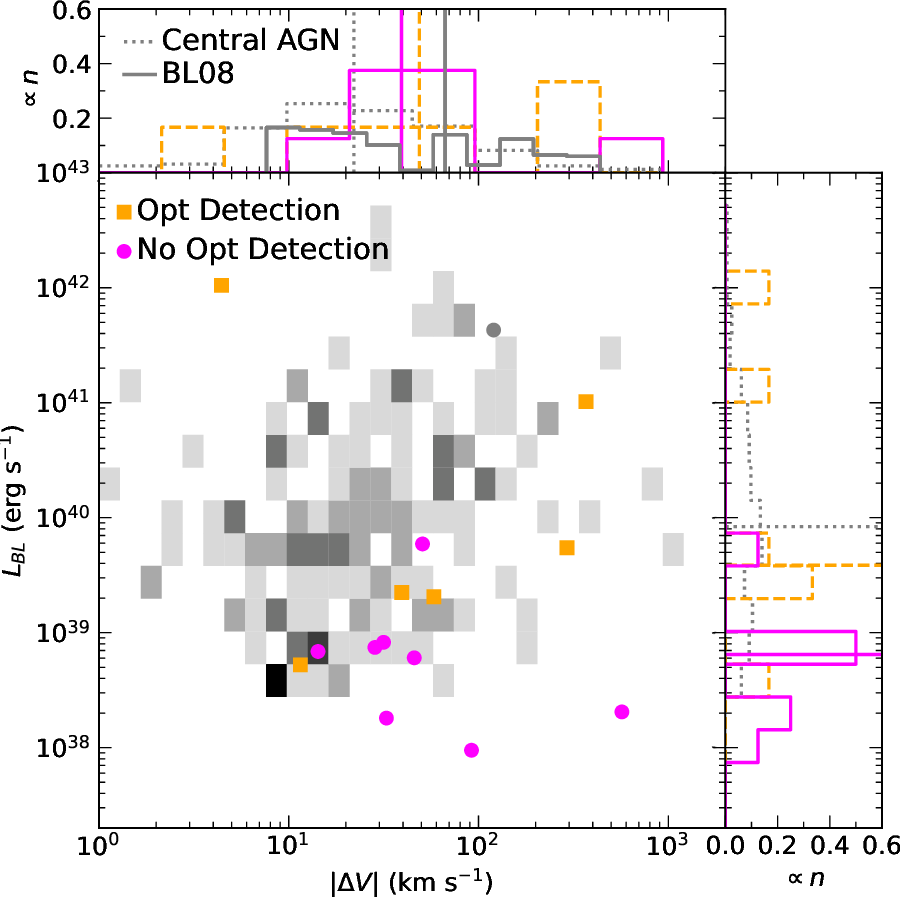}
\caption{\footnotesize{Broad emission line luminosity (\LBL; total value from all spaxels) against the magnitude of the projected velocity offset relative to the host galaxy systemic velocity (\DeltaV; flux-weighted average from all spaxels) for the central BLRs (linear-scale density map) and the subset of spatially offset BLRs with (orange squares) and without (magenta circles) optical image counterpart detections. The distributions of \LBL~and \DeltaV~are shown along the right and top axes, respectively, for the offset BLRs with (orange dashed) and without (magenta solid) image counterpart detections and for the central BLRs (gray dotted). The mean \DeltaVMag~values of the spatially offset samples show a bias toward larger values relative to the central sample (by \MeanOffPercrpDeltaVvoffwt\%). For comparison, the simulated distribution of recoiling MBH projected velocities from \citet{Blecha:2008} (BL08) is also shown (solid gray line), and the offset BLR mean values are consistent with it to within \MeanOffPercrbDeltaVvoffwt\%.
}}
\label{fig:LLINE_DELTAV}
\end{figure}

The left panel of Figure \ref{fig:LLINE_DELTAV} shows the total BLR luminosities (\LBL; integrated over all spaxels; Table \ref{tab:Samp}) against the magnitudes of the BLR projected velocity offsets from the host galaxy systemic (\DeltaV; flux-weighted average of the velocity offsets in the individual spaxels; Table \ref{tab:Samp}). The offset AGN samples with and without optical counterparts have similar \DeltaVMag~values (means within 10\,\uV~of each other) and their distributions overlap significantly with that of the central AGN. However, the offset AGN samples do show evidence for systematically larger mean \DeltaVMag~values (by \MeanOffPercrpDeltaVvoffwt\%). Though these values are all within the range typically expected for galaxy rotation \citep[e.g.,][]{Sofue:2001}, this result may be indicative of motion of the spatially offset BLRs relative to the host galaxies, and the mean \DeltaVMag~value is consistent (within \MeanOffPercrbDeltaVvoffwt\%) with the predicted observed values for recoils in \citet{Blecha:2008}.

Furthermore, the relatively lower \LBL~values for the offset AGN without optical image counterparts (Section \ref{sec:BPT}) may be consistent with the recoil scenario in that the kick velocities are expected to reduce the AGN duty cycles since they are displaced from large reservoirs of nuclear gas and dust for accretion \citep{Blecha:2016}. However, contrary to predictions that observable recoils will be preferentially hosted by galaxies with low escape velocities, a bias toward lower host galaxy stellar masses is not observed.

\begin{figure}[ht!]
\includegraphics[width=0.47\textwidth]{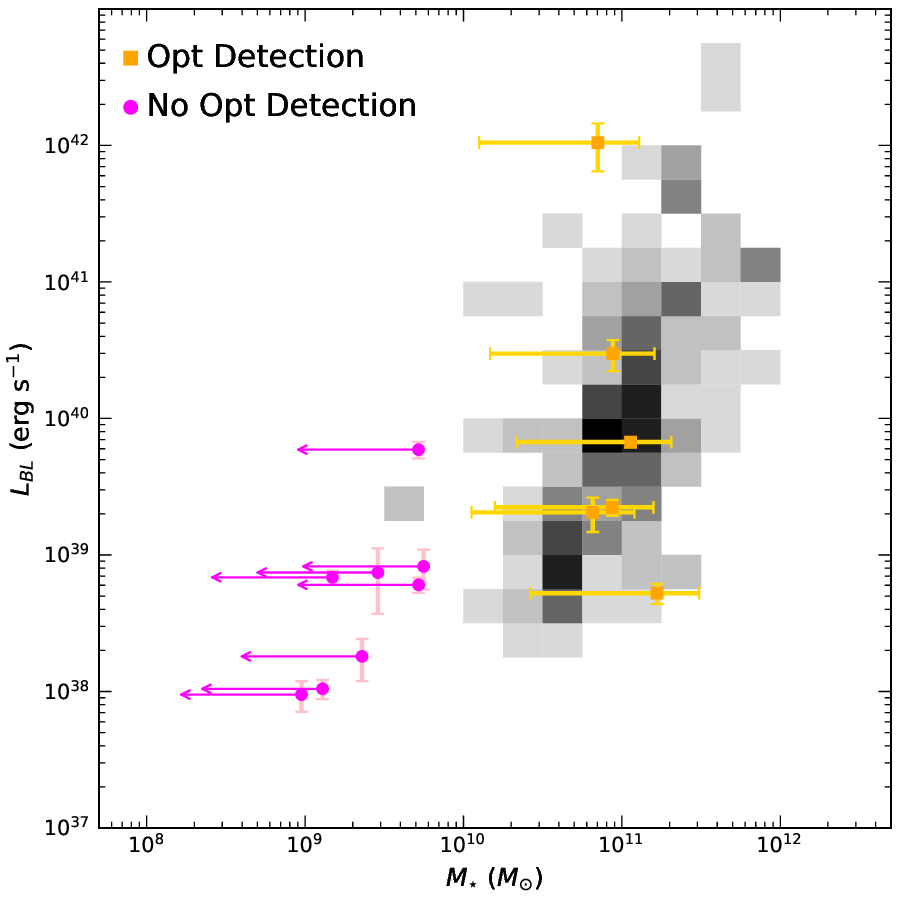}
\caption{\footnotesize{Broad emission line luminosities (\LBL) against the optical image counterpart stellar masses (\Mstar; orange squares) or upper limits (for those without counterpart detections; magenta circles and left-pointing arrows). The density map shows the central AGN sample from this work. The subset without counterpart detections has relatively large \LBL/\Mstar~ratios compared to central AGN, potentially due to merger-driven AGN enhancements and/or tidal stripping. Those without counterpart detections are also consistent with being ejected MBHs within faint hypercompact stellar systems.}}
\label{fig:LLINE_MSTAR}
\end{figure}

\section{Results}
\label{sec:results}

In the following sections, we explore the implications of this offset BLR sample for the role of galaxy mergers in AGN triggering (Section \ref{sec:accretion}), dual AGN and photo-ionization of the ISM on extended spatial scales (Section \ref{sec:extended}), and the frequency of recoiling AGN detections (Section \ref{sec:density}).

\subsection{Merger-Driven AGN Accretion}
\label{sec:accretion}

While galaxy mergers are effective at triggering accretion onto MBHs, whether or not this process plays a significant role in the overall AGN population is unclear. In particular, several numerical analyses suggest that the fraction of galaxies with AGN is enhanced for those in on-going mergers or merger remnants \citep[e.g.,][]{McAlpine:2020,Byrne-Mamahit:2022}. However, alternative results instead predict no enhancements and that the AGN luminosities are primarily regulated by the masses of their host galaxy stellar cores \citep[e.g.,][]{Steinborn:2016,Weigel:2018}.

The spatially constrained positions of offset AGN allow for masses or upper limits of their stellar cores to be measured, thereby facilitating tests of these predictions. Furthermore, while selection of AGN in galaxy mergers is often based on morphological signatures with clear disturbances that preferentially find major mergers \citep[e.g.,][]{Cisternas:2011,Kocevski:2012,Villforth:2014}, the selection in this work is based on AGN signatures and is therefore also sensitive to the population of minor mergers with extreme mass ratios that dominate merger rates \citep[e.g.,][]{Lotz:2011,Duncan:2019}.

Figure \ref{fig:LLINE_MSTAR} shows the relationship between \LBL~and the stellar core mass (or upper limit) for the central AGN and the offset AGN (\Mstar). The overall sample does not show a bias toward relatively high \LBL/\Mstar~values compared to central AGN (consistent within 1$\sigma$). If all of the offset AGN in this work (including those without optical image counterparts) are infalling AGN, this result would indicate that MBH accretion in these systems is relatively independent of the merger process. However, previous work suggests that merger-driven MBH growth may be strongest for AGN in relatively low mass galaxies \citep[e.g.,][]{Yu:2011,Capelo:2015,Comerford:2015,Barrows:2023}. Indeed, when limiting our sample to the subset without stellar core detections, the median value of \LBL/\Mstar~is $\ge$\,\LLineMstarRatioDiff~times larger than that of the central AGN sample.

\subsection{Dual AGN and Extended Photo-ionization}
\label{sec:extended}

A fraction of galaxy mergers are observed to host AGN in the nuclei of both galaxies \citep[dual AGN; e.g.,][]{Komossa2003,Bianchi2008,Comerford2009b,Piconcelli2010,Koss:2012,Mazzarella:2012,Comerford:2015,Mueller-Sanchez:2015,Hou:2022}. The majority of AGN in such systems are buried under obscuring material that may also fuel the strongest stages of merger-driven MBH growth \citep[e.g.,][]{Blecha:2018}. Optically-selected dual AGN in the nearby Universe, on the other hand, may represent sources in a relatively late stage of merger-driven AGN evolution after radiation pressure has removed obscuring material \citep[e.g.,][]{Hopkins2008,Hickox:2009}, thereby allowing tests of this connection.

Figure \ref{fig:F_AGN_bl_nuc} shows that the \FAN~values are \PercDiffnucrofas\% higher for the offset broad line AGN with optical image counterparts relative to those without. Furthermore, the KS statistic and null hypothesis probability are \KSStatnucrofan~and \KSPnucrofan, respectively (for both the \nii~and \sii~diagnostics) that they have similar \FAN~distributions. This observation may suggest that the offset broad line AGN in more massive galaxies have a higher probability of hosting dual AGN (i.e., a second optically-selected AGN in the host galaxy nucleus). This preference may be explained by their larger optical image counterpart stellar masses, since more massive galaxies are more likely to be in major mergers \citep[e.g.,][]{Weigel:2018}, and major mergers may be more likely to host AGN \citep[e.g.,][]{Somerville:2001,Cox:2008,Ellison:2008}. Hence, despite the lack of significant obscuring material, correlated triggering of optically-selected AGN in mergers may be efficient \citep[e.g.,][]{Fu:2018}.

Alternatively, if no central AGN exists, these observations may be explained by the relatively higher \LBL~values of the offset AGN with optical image counterparts (i.e., Figure \ref{fig:LLINE_DELTAV}) that may photo-ionize the ISM on spatially extended scales, including around the host galaxy nucleus. Indeed, 3 of the \combOptSZ~likely infalling AGN have \FABL\,$>$\,0 (between the \nii~and \sii~diagnostics combined) that suggest the AGN emission is strong enough to dominate the ionization potential over that of star formation. This possibility has implications for the nature of host galaxies with BPT indications of AGN, suggesting that not only can such narrow emission line ratios be produced by off-nuclear AGN, but they can also mimic the appearance of a nuclear BPT-selected AGN.

\subsection{Surface Density of Recoiling AGN}
\label{sec:density}

Based on the survey footprint of \ma, the offset BLR sample without optical image counterparts from this work implies \SurfDens~recoiling AGN candidates per deg$^{2}$ with visible BLRs (an upper limit to the true surface density considering they are candidates). For comparison, assuming the random spin alignment model from \citet{Blecha:2016}, this is higher (by up to $\sim$\,1\,dex) than the predictions for photometric detections of recoiling AGN for all surveys considered in that work at $z$\,$<$\,0.15 (approximate maximum redshift for the final \ma~sample). Hence, since our measured surface density is an upper limit, it is consistent with these predictions and does not suggest highly efficient spin alignment in MBH binaries.

\section{Conclusions}
\label{sec:conc}

From the spatially-resolved spectroscopy of galaxies in the \ma~survey, we have developed a sample of \combOffSZ~candidate broad line AGN that are spatially offset from their host galaxy centroid by $>$\,3 times the standard deviation of the source extent and by more than the \ma~FWHM spatial resolution of 2\farcs5 (\combOffSZ~and 1 from \ha~and \hb~detections, respectively, with an overlap of 1). The sample is based on the catalog of galaxies with broad emission lines (line widths of FWHM\,$>$\,1000\,\uV) from \citet{Negus:2024}, which itself was derived from the 10,010 unique galaxies in the final \ma~survey. This is the largest uniform sample of spectroscopically detected spatially offset broad line AGN candidates and the first such systematically generated sample from an IFS survey. The primary motivation for this project is to identify candidate recoiling or slingshot MBHs since such sources, if accreting as AGN, are expected to have unobscured BLRs that may have measurable nuclear offsets. Below we synthesize our interpretation of the nature of this sample and implications for merger-driven MBH growth and detection of recoiling AGN:

\begin{enumerate}

\item Regarding explanations that do not invoke offset AGN, we conclude the following: A) the probability of background or foreground chance projections is $<2.7\times10^{-4}$; B) based on \oiii~emission line models, no evidence is seen for narrow line outflows strong enough to mimic the presence of broad AGN components; C) while a supernova origin for the broad emission lines is expected to be rare, it can not be ruled out for any individual sources.

\item Based on the available SDSS imaging, optical counterparts are identified for \combOptPerc\% (\combOptSZ/\combOffSZ) of the sample. The stellar masses associated with the optical counterpart detections significantly exceed predictions for bound hypercompact stellar systems around recoiling MBHs. Therefore, they are likely to be infalling AGN in on-going galaxy mergers. The remaining \combNoOptPerc\% (\combNoOptSZ/\combOffSZ) only have upper limit masses on stellar counterparts and are therefore potentially consistent with being recoiling or slingshot MBHs.

\item Based on BPT diagnostics, the AGN-ionized narrow emission line signatures for the subset of the BLRs without optical image counterparts are \PercDiffblrofas\% weaker than those for the subset with counterpart detections. An absence of narrow line AGN emission at the BLR position is consistent with predictions for the recoil or slingshot scenarios.

\item The projected velocity offsets (relative to the host galaxy systemic) of the BLR candidates that are spatially offset are \MeanOffPercrpDeltaVvoffwt\% larger than for those in galaxy nuclei. This result suggests they may represent motion of the BLRs relative to the host galaxy, and they are also consistent (to within \MeanOffPercrbDeltaVvoffwt\%) with predictions for observed recoiling AGN velocities. Furthermore, those without optical image counterparts have relatively lower luminosities (by a factor of \LLineRatioDiffInt), qualitatively consistent with predictions for recoiling AGN to accrete less efficiently.

\item In the scenario that the recoil/slingshot AGN candidates are actually infalling MBHs, those without optical counterpart detections have relatively large broad emission line luminosity to stellar counterpart mass ratios ($>$\,\LLineMstarRatioDiff~times larger than for the central AGN). This result would be consistent with merger-driven triggering of broad line AGN being stronger in relatively low mass galaxies.

\item The AGN-ionized narrow emission line signatures in the host galaxy nuclei of the offset AGN candidates are \PercDiffnucrofas\% stronger for those with optical image counterpart detections compared to those without counterparts. This may indicate they are more likely to host dual AGN, possibly since they are in more equal mass mergers with stronger tidal forces on each galaxy. Alternatively, since the offset AGN candidates with optical image counterparts are more luminous, this result may indicate that they are photo-ionizing the ISM on spatially extended scales.

\item The surface density of ejected AGN candidates from this work is up to $\sim$\,1\,dex higher than predictions for recoiling AGN following random MBH spin alignments that can be found in photometric surveys. Since this sample can only provide an upper limit, this result is consistent with these predictions and hence does not suggest highly efficient MBH spin alignment in binaries.

\end{enumerate}

In summary, the sample developed through this work contains \combNoOptSZ~sources that are plausible candidates for ejected (recoil or slingshot) MBHs. Circumstantial evidence in favor of the ejected MBH scenario is in the form of BLRs with systematic velocity offsets from the host galaxy that are consistent with predictions for recoils, relatively low AGN luminosities, and relatively weak NLR AGN emission at the location of the BLRs. Those without evidence for NLR AGN emission in the host galaxy nucleus (J0946, J1124, J1508, J1604, and J1611) may be the strongest recoil candidates given the longer timescales for predicted recoil observability compared to those of AGN light echos. The remainder may be stronger candidates for slingshot MBHs. However, deeper imaging to search for faint stellar cores associated with the candidates is necessary for further consideration of the recoil or slingshot scenarios.

\begin{acknowledgments}

We thank an anonymous reviewer for the detailed and thorough comments that have improved the manuscript quality. The authors thank Laura Blecha for insightful discussions that contributed to the interpretation of the results. J.M.C. and J.N. acknowledge support from NSF AST1714503 and NSF AST1847938. F.M-S. acknowledges support from NASA through awards 80NSSC19K1096 and 80NSSC23K1529. This project makes use of the MaNGA-Pipe3D dataproducts. We thank the IA-UNAM MaNGA team for creating this catalogue, and the Conacyt Project CB-285080 for supporting them. Funding for the Sloan Digital Sky Survey IV has been provided by the Alfred P. Sloan Foundation, the U.S.  Department of Energy Office of Science, and the Participating Institutions. SDSS-IV acknowledges support and resources from the Center for High Performance Computing  at the University of Utah. The SDSS website is www.sdss4.org. SDSS-IV is managed by the Astrophysical Research Consortium for the Participating Institutions of the SDSS Collaboration including the Brazilian Participation Group, the Carnegie Institution for Science, Carnegie Mellon University, Center for Astrophysics | Harvard \& Smithsonian, the Chilean Participation Group, the French Participation Group, Instituto de Astrof\'isica de Canarias, The Johns Hopkins University, Kavli Institute for the Physics and Mathematics of the Universe (IPMU) / University of Tokyo, the Korean Participation Group, Lawrence Berkeley National Laboratory, Leibniz Institut f\"ur Astrophysik Potsdam (AIP),  Max-Planck-Institut f\"ur Astronomie (MPIA Heidelberg), Max-Planck-Institut f\"ur Astrophysik (MPA Garching), Max-Planck-Institut f\"ur Extraterrestrische Physik (MPE), National Astronomical Observatories of China, New Mexico State University, New York University, University of Notre Dame, Observat\'ario Nacional / MCTI, The Ohio State University, Pennsylvania State University, Shanghai Astronomical Observatory, United Kingdom Participation Group, Universidad Nacional Aut\'onoma de M\'exico, University of Arizona, University of Colorado Boulder, University of Oxford, University of Portsmouth, University of Utah, University of Virginia, University of Washington, University of Wisconsin, Vanderbilt University, and Yale University.

\end{acknowledgments}

\facilities{Sloan}

\software{\astropy\footnote{\href{\astropylink}{\astropylink}} \citep{astropy:2013, astropy:2018,astropy:2022}, \marvin\footnote{\href{\marvinlink}{\marvinlink}} \citep{Cherinka:2019}.}

\newpage

\appendix

\section{Best-Fitting Spectral Models}
\label{sec:spectral_models}

\begin{figure*}[ht!]
\includegraphics[width=0.98\textwidth]{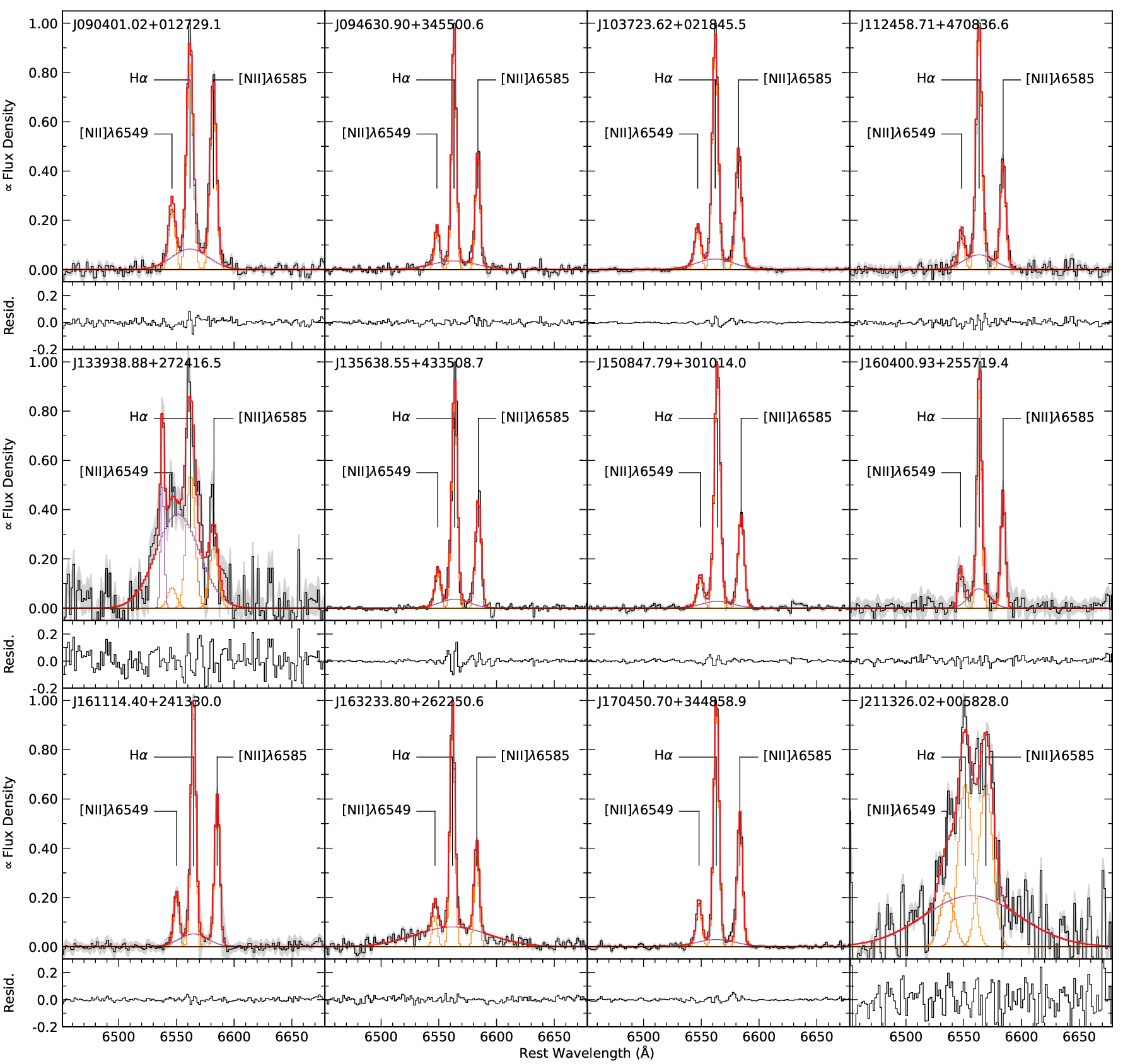}
\caption{\footnotesize{Same as Figure \ref{fig:spectra_0}  but for the remainder of the sample (sorted by right ascension).}}
\label{fig:spectra_1}
\end{figure*}

Figure \ref{fig:spectra_1} shows the spectral regions and best-fit models for which broad emission line components are detected (see Figure \ref{fig:spectra_0} for the others). The spectral models and broad emission line detections are from \citet{Negus:2024}. The blue excess peak in J133938.88+272416.5 resembles those of double-peaked broad line emitters and\textbf{ has been fit with an additional Gaussian component.}

\begin{thebibliography}{}
\expandafter\ifx\csname natexlab\endcsname\relax\def\natexlab#1{#1}\fi
\providecommand{\url}[1]{\href{#1}{#1}}
\providecommand{\dodoi}[1]{doi:~\href{http://doi.org/#1}{\nolinkurl{#1}}}
\providecommand{\doeprint}[1]{\href{http://ascl.net/#1}{\nolinkurl{http://ascl.net/#1}}}
\providecommand{\doarXiv}[1]{\href{https://arxiv.org/abs/#1}{\nolinkurl{https://arxiv.org/abs/#1}}}

\bibitem[{{Astropy Collaboration} {et~al.}(2013){Astropy Collaboration},
  {Robitaille}, {Tollerud}, {Greenfield}, {Droettboom}, {Bray}, {Aldcroft},
  {Davis}, {Ginsburg}, {Price-Whelan}, {Kerzendorf}, {Conley}, {Crighton},
  {Barbary}, {Muna}, {Ferguson}, {Grollier}, {Parikh}, {Nair}, {Unther},
  {Deil}, {Woillez}, {Conseil}, {Kramer}, {Turner}, {Singer}, {Fox}, {Weaver},
  {Zabalza}, {Edwards}, {Azalee Bostroem}, {Burke}, {Casey}, {Crawford},
  {Dencheva}, {Ely}, {Jenness}, {Labrie}, {Lim}, {Pierfederici}, {Pontzen},
  {Ptak}, {Refsdal}, {Servillat}, \& {Streicher}}]{astropy:2013}
{Astropy Collaboration}, {Robitaille}, T.~P., {Tollerud}, E.~J., {et~al.} 2013,
  \aap, 558, A33, \dodoi{10.1051/0004-6361/201322068}

\bibitem[{{Astropy Collaboration} {et~al.}(2018){Astropy Collaboration},
  {Price-Whelan}, {Sip{\H o}cz}, {G{\"u}nther}, {Lim}, {Crawford}, {Conseil},
  {Shupe}, {Craig}, {Dencheva}, {Ginsburg}, {VanderPlas}, {Bradley},
  {P{\'e}rez-Su{\'a}rez}, {de Val-Borro}, {Aldcroft}, {Cruz}, {Robitaille},
  {Tollerud}, {Ardelean}, {Babej}, {Bach}, {Bachetti}, {Bakanov}, {Bamford},
  {Barentsen}, {Barmby}, {Baumbach}, {Berry}, {Biscani}, {Boquien}, {Bostroem},
  {Bouma}, {Brammer}, {Bray}, {Breytenbach}, {Buddelmeijer}, {Burke},
  {Calderone}, {Cano Rodr{\'{\i}}guez}, {Cara}, {Cardoso}, {Cheedella},
  {Copin}, {Corrales}, {Crichton}, {D'Avella}, {Deil}, {Depagne}, {Dietrich},
  {Donath}, {Droettboom}, {Earl}, {Erben}, {Fabbro}, {Ferreira}, {Finethy},
  {Fox}, {Garrison}, {Gibbons}, {Goldstein}, {Gommers}, {Greco}, {Greenfield},
  {Groener}, {Grollier}, {Hagen}, {Hirst}, {Homeier}, {Horton}, {Hosseinzadeh},
  {Hu}, {Hunkeler}, {Ivezi{\'c}}, {Jain}, {Jenness}, {Kanarek}, {Kendrew},
  {Kern}, {Kerzendorf}, {Khvalko}, {King}, {Kirkby}, {Kulkarni}, {Kumar},
  {Lee}, {Lenz}, {Littlefair}, {Ma}, {Macleod}, {Mastropietro}, {McCully},
  {Montagnac}, {Morris}, {Mueller}, {Mumford}, {Muna}, {Murphy}, {Nelson},
  {Nguyen}, {Ninan}, {N{\"o}the}, {Ogaz}, {Oh}, {Parejko}, {Parley}, {Pascual},
  {Patil}, {Patil}, {Plunkett}, {Prochaska}, {Rastogi}, {Reddy Janga},
  {Sabater}, {Sakurikar}, {Seifert}, {Sherbert}, {Sherwood-Taylor}, {Shih},
  {Sick}, {Silbiger}, {Singanamalla}, {Singer}, {Sladen}, {Sooley},
  {Sornarajah}, {Streicher}, {Teuben}, {Thomas}, {Tremblay}, {Turner},
  {Terr{\'o}n}, {van Kerkwijk}, {de la Vega}, {Watkins}, {Weaver}, {Whitmore},
  {Woillez}, {Zabalza}, \& {Astropy Contributors}}]{astropy:2018}
{Astropy Collaboration}, {Price-Whelan}, A.~M., {Sip{\H o}cz}, B.~M., {et~al.}
  2018, \aj, 156, 123, \dodoi{10.3847/1538-3881/aabc4f}

\bibitem[{{Astropy Collaboration} {et~al.}(2022){Astropy Collaboration},
  {Price-Whelan}, {Lim}, {Earl}, {Starkman}, {Bradley}, {Shupe}, {Patil},
  {Corrales}, {Brasseur}, {N{\"o}the}, {Donath}, {Tollerud}, {Morris},
  {Ginsburg}, {Vaher}, {Weaver}, {Tocknell}, {Jamieson}, {van Kerkwijk},
  {Robitaille}, {Merry}, {Bachetti}, {G{\"u}nther}, {Aldcroft},
  {Alvarado-Montes}, {Archibald}, {B{\'o}di}, {Bapat}, {Barentsen},
  {Baz{\'a}n}, {Biswas}, {Boquien}, {Burke}, {Cara}, {Cara}, {Conroy},
  {Conseil}, {Craig}, {Cross}, {Cruz}, {D'Eugenio}, {Dencheva}, {Devillepoix},
  {Dietrich}, {Eigenbrot}, {Erben}, {Ferreira}, {Foreman-Mackey}, {Fox},
  {Freij}, {Garg}, {Geda}, {Glattly}, {Gondhalekar}, {Gordon}, {Grant},
  {Greenfield}, {Groener}, {Guest}, {Gurovich}, {Handberg}, {Hart},
  {Hatfield-Dodds}, {Homeier}, {Hosseinzadeh}, {Jenness}, {Jones}, {Joseph},
  {Kalmbach}, {Karamehmetoglu}, {Ka{\l}uszy{\'n}ski}, {Kelley}, {Kern},
  {Kerzendorf}, {Koch}, {Kulumani}, {Lee}, {Ly}, {Ma}, {MacBride}, {Maljaars},
  {Muna}, {Murphy}, {Norman}, {O'Steen}, {Oman}, {Pacifici}, {Pascual},
  {Pascual-Granado}, {Patil}, {Perren}, {Pickering}, {Rastogi}, {Roulston},
  {Ryan}, {Rykoff}, {Sabater}, {Sakurikar}, {Salgado}, {Sanghi}, {Saunders},
  {Savchenko}, {Schwardt}, {Seifert-Eckert}, {Shih}, {Jain}, {Shukla}, {Sick},
  {Simpson}, {Singanamalla}, {Singer}, {Singhal}, {Sinha}, {Sip{\H{o}}cz},
  {Spitler}, {Stansby}, {Streicher}, {{\v{S}}umak}, {Swinbank}, {Taranu},
  {Tewary}, {Tremblay}, {de Val-Borro}, {Van Kooten}, {Vasovi{\'c}}, {Verma},
  {de Miranda Cardoso}, {Williams}, {Wilson}, {Winkel}, {Wood-Vasey}, {Xue},
  {Yoachim}, {Zhang}, {Zonca}, \& {Astropy Project
  Contributors}}]{astropy:2022}
{Astropy Collaboration}, {Price-Whelan}, A.~M., {Lim}, P.~L., {et~al.} 2022,
  \apj, 935, 167, \dodoi{10.3847/1538-4357/ac7c74}

\bibitem[{{Barrows} {et~al.}(2018){Barrows}, {Comerford}, \&
  {Greene}}]{Barrows:2018}
{Barrows}, R.~S., {Comerford}, J.~M., \& {Greene}, J.~E. 2018, \apj, 869, 154,
  \dodoi{10.3847/1538-4357/aaedb6}

\bibitem[{{Barrows} {et~al.}(2016){Barrows}, {Comerford}, {Greene}, \&
  {Pooley}}]{Barrows:2016}
{Barrows}, R.~S., {Comerford}, J.~M., {Greene}, J.~E., \& {Pooley}, D. 2016,
  \apj, 829, 37, \dodoi{10.3847/0004-637X/829/1/37}

\bibitem[{{Barrows} {et~al.}(2017){Barrows}, {Comerford}, {Greene}, \&
  {Pooley}}]{Barrows:2017}
---. 2017, \apj, 838, 129, \dodoi{10.3847/1538-4357/aa64d9}

\bibitem[{{Barrows} {et~al.}(2023){Barrows}, {Comerford}, {Stern}, \&
  {Assef}}]{Barrows:2023}
{Barrows}, R.~S., {Comerford}, J.~M., {Stern}, D., \& {Assef}, R.~J. 2023,
  \apj, 951, 92, \dodoi{10.3847/1538-4357/acd2d3}

\bibitem[{{Barrows} {et~al.}(2011){Barrows}, {Lacy}, {Kennefick}, {Kennefick},
  \& {Seigar}}]{Barrows:2011}
{Barrows}, R.~S., {Lacy}, C.~H.~S., {Kennefick}, D., {Kennefick}, J., \&
  {Seigar}, M.~S. 2011, \na, 16, 122, \dodoi{10.1016/j.newast.2010.08.004}

\bibitem[{{Barrows} {et~al.}(2024){Barrows}, {Mezcua}, {Comerford}, \&
  {Stern}}]{Barrows:2024}
{Barrows}, R.~S., {Mezcua}, M., {Comerford}, J.~M., \& {Stern}, D. 2024, \apj,
  964, 187, \dodoi{10.3847/1538-4357/ad25fe}

\bibitem[{{Bekenstein}(1973)}]{Bekenstein:1973}
{Bekenstein}, J.~D. 1973, \apj, 183, 657, \dodoi{10.1086/152255}

\bibitem[{{Bertin} \& {Arnouts}(1996)}]{Bertin:Arnouts:1996}
{Bertin}, E., \& {Arnouts}, S. 1996, \aaps, 117, 393

\bibitem[{{Bianchi} {et~al.}(2008){Bianchi}, {Chiaberge}, {Piconcelli},
  {Guainazzi}, \& {Matt}}]{Bianchi2008}
{Bianchi}, S., {Chiaberge}, M., {Piconcelli}, E., {Guainazzi}, M., \& {Matt},
  G. 2008, \mnras, 386, 105, \dodoi{10.1111/j.1365-2966.2008.13078.x}

\bibitem[{{Blecha} {et~al.}(2019){Blecha}, {Brisken}, {Burke-Spolaor},
  {Civano}, {Comerford}, {Darling}, {Lazio}, \& {Maccarone}}]{Blecha:2019}
{Blecha}, L., {Brisken}, W., {Burke-Spolaor}, S., {et~al.} 2019, Astro2020:
  Decadal Survey on Astronomy and Astrophysics, 2020, 318,
  \dodoi{10.48550/arXiv.1903.09301}

\bibitem[{{Blecha} {et~al.}(2013){Blecha}, {Civano}, {Elvis}, \&
  {Loeb}}]{Blecha:2013b}
{Blecha}, L., {Civano}, F., {Elvis}, M., \& {Loeb}, A. 2013, \mnras, 428, 1341,
  \dodoi{10.1093/mnras/sts114}

\bibitem[{{Blecha} {et~al.}(2011){Blecha}, {Cox}, {Loeb}, \&
  {Hernquist}}]{Blecha:2011}
{Blecha}, L., {Cox}, T.~J., {Loeb}, A., \& {Hernquist}, L. 2011, \mnras, 412,
  2154, \dodoi{10.1111/j.1365-2966.2010.18042.x}

\bibitem[{{Blecha} \& {Loeb}(2008)}]{Blecha:2008}
{Blecha}, L., \& {Loeb}, A. 2008, \mnras, 390, 1311,
  \dodoi{10.1111/j.1365-2966.2008.13790.x}

\bibitem[{{Blecha} {et~al.}(2018){Blecha}, {Snyder}, {Satyapal}, \&
  {Ellison}}]{Blecha:2018}
{Blecha}, L., {Snyder}, G.~F., {Satyapal}, S., \& {Ellison}, S.~L. 2018,
  \mnras, 478, 3056, \dodoi{10.1093/mnras/sty1274}

\bibitem[{{Blecha} {et~al.}(2016){Blecha}, {Sijacki}, {Kelley}, {Torrey},
  {Vogelsberger}, {Nelson}, {Springel}, {Snyder}, \& {Hernquist}}]{Blecha:2016}
{Blecha}, L., {Sijacki}, D., {Kelley}, L.~Z., {et~al.} 2016, \mnras, 456, 961,
  \dodoi{10.1093/mnras/stv2646}

\bibitem[{{Bogdanovi{\'c}} {et~al.}(2009{\natexlab{a}}){Bogdanovi{\'c}},
  {Eracleous}, \& {Sigurdsson}}]{Bogdanovic09a}
{Bogdanovi{\'c}}, T., {Eracleous}, M., \& {Sigurdsson}, S. 2009{\natexlab{a}},
  \na, 53, 113, \dodoi{10.1016/j.newar.2009.09.005}

\bibitem[{{Bogdanovi{\'c}} {et~al.}(2009{\natexlab{b}}){Bogdanovi{\'c}},
  {Eracleous}, \& {Sigurdsson}}]{Bogdanovic09b}
---. 2009{\natexlab{b}}, ApJ, 697, 288, \dodoi{10.1088/0004-637X/697/1/288}

\bibitem[{{Bonetti} {et~al.}(2016){Bonetti}, {Haardt}, {Sesana}, \&
  {Barausse}}]{Bonetti:2016}
{Bonetti}, M., {Haardt}, F., {Sesana}, A., \& {Barausse}, E. 2016, \mnras, 461,
  4419, \dodoi{10.1093/mnras/stw1590}

\bibitem[{{Bonetti} {et~al.}(2018){Bonetti}, {Haardt}, {Sesana}, \&
  {Barausse}}]{Bonetti:2018}
---. 2018, \mnras, 477, 3910, \dodoi{10.1093/mnras/sty896}

\bibitem[{{Bonning} {et~al.}(2007){Bonning}, {Shields}, \&
  {Salviander}}]{Bonning:2007}
{Bonning}, E.~W., {Shields}, G.~A., \& {Salviander}, S. 2007, \apjl, 666, L13,
  \dodoi{10.1086/521674}

\bibitem[{{Boroson} \& {Lauer}(2009)}]{BL09}
{Boroson}, T.~A., \& {Lauer}, T.~R. 2009, Nature, 458, 53,
  \dodoi{10.1038/nature07779}

\bibitem[{{Bundy} {et~al.}(2015){Bundy}, {Bershady}, {Law}, {Yan}, {Drory},
  {MacDonald}, {Wake}, {Cherinka}, {S{\'a}nchez-Gallego}, {Weijmans}, {Thomas},
  {Tremonti}, {Masters}, {Coccato}, {Diamond-Stanic}, {Arag{\'o}n-Salamanca},
  {Avila-Reese}, {Badenes}, {Falc{\'o}n-Barroso}, {Belfiore}, {Bizyaev},
  {Blanc}, {Bland-Hawthorn}, {Blanton}, {Brownstein}, {Byler}, {Cappellari},
  {Conroy}, {Dutton}, {Emsellem}, {Etherington}, {Frinchaboy}, {Fu}, {Gunn},
  {Harding}, {Johnston}, {Kauffmann}, {Kinemuchi}, {Klaene}, {Knapen},
  {Leauthaud}, {Li}, {Lin}, {Maiolino}, {Malanushenko}, {Malanushenko}, {Mao},
  {Maraston}, {McDermid}, {Merrifield}, {Nichol}, {Oravetz}, {Pan}, {Parejko},
  {Sanchez}, {Schlegel}, {Simmons}, {Steele}, {Steinmetz}, {Thanjavur},
  {Thompson}, {Tinker}, {van den Bosch}, {Westfall}, {Wilkinson}, {Wright},
  {Xiao}, \& {Zhang}}]{Bundy:2015}
{Bundy}, K., {Bershady}, M.~A., {Law}, D.~R., {et~al.} 2015, \apj, 798, 7,
  \dodoi{10.1088/0004-637X/798/1/7}

\bibitem[{{Byrne-Mamahit} {et~al.}(2022){Byrne-Mamahit}, {Hani}, {Ellison},
  {Quai}, \& {Patton}}]{Byrne-Mamahit:2022}
{Byrne-Mamahit}, S., {Hani}, M., {Ellison}, S., {Quai}, S., \& {Patton}, D.
  2022, arXiv e-prints, arXiv:2212.07342.
\newblock \doarXiv{2212.07342}

\bibitem[{{Campanelli} {et~al.}(2007{\natexlab{a}}){Campanelli}, {Lousto},
  {Zlochower}, \& {Merritt}}]{Campanelli:2007a}
{Campanelli}, M., {Lousto}, C., {Zlochower}, Y., \& {Merritt}, D.
  2007{\natexlab{a}}, \apjl, 659, L5, \dodoi{10.1086/516712}

\bibitem[{{Campanelli} {et~al.}(2007{\natexlab{b}}){Campanelli}, {Lousto},
  {Zlochower}, \& {Merritt}}]{Campanelli:2007b}
{Campanelli}, M., {Lousto}, C.~O., {Zlochower}, Y., \& {Merritt}, D.
  2007{\natexlab{b}}, \prl, 98, 231102, \dodoi{10.1103/PhysRevLett.98.231102}

\bibitem[{{Capelo} {et~al.}(2015){Capelo}, {Volonteri}, {Dotti}, {Bellovary},
  {Mayer}, \& {Governato}}]{Capelo:2015}
{Capelo}, P.~R., {Volonteri}, M., {Dotti}, M., {et~al.} 2015, \mnras, 447,
  2123, \dodoi{10.1093/mnras/stu2500}

\bibitem[{{Centrella} {et~al.}(2010){Centrella}, {Baker}, {Kelly}, \& {van
  Meter}}]{Centrella:2010}
{Centrella}, J., {Baker}, J.~G., {Kelly}, B.~J., \& {van Meter}, J.~R. 2010,
  Reviews of Modern Physics, 82, 3069, \dodoi{10.1103/RevModPhys.82.3069}

\bibitem[{{Cherinka} {et~al.}(2019){Cherinka}, {Andrews},
  {S{\'a}nchez-Gallego}, {Brownstein}, {Argudo-Fern{\'a}ndez}, {Blanton},
  {Bundy}, {Jones}, {Masters}, {Law}, {Rowlands}, {Weijmans}, {Westfall}, \&
  {Yan}}]{Cherinka:2019}
{Cherinka}, B., {Andrews}, B.~H., {S{\'a}nchez-Gallego}, J., {et~al.} 2019,
  \aj, 158, 74, \dodoi{10.3847/1538-3881/ab2634}

\bibitem[{{Chiaberge} {et~al.}(2018){Chiaberge}, {Tremblay}, {Capetti}, \&
  {Norman}}]{Chiaberge:2018}
{Chiaberge}, M., {Tremblay}, G.~R., {Capetti}, A., \& {Norman}, C. 2018, \apj,
  861, 56, \dodoi{10.3847/1538-4357/aac48b}

\bibitem[{{Choksi} {et~al.}(2017){Choksi}, {Behroozi}, {Volonteri},
  {Schneider}, {Ma}, {Silk}, \& {Moster}}]{Choksi:2017}
{Choksi}, N., {Behroozi}, P., {Volonteri}, M., {et~al.} 2017, \mnras, 472,
  1526, \dodoi{10.1093/mnras/stx2113}

\bibitem[{{Cisternas} {et~al.}(2011){Cisternas}, {Jahnke}, {Inskip},
  {Kartaltepe}, {Koekemoer}, {Lisker}, {Robaina}, {Scodeggio}, {Sheth},
  {Trump}, {Andrae}, {Miyaji}, {Lusso}, {Brusa}, {Capak}, {Cappelluti},
  {Civano}, {Ilbert}, {Impey}, {Leauthaud}, {Lilly}, {Salvato}, {Scoville}, \&
  {Taniguchi}}]{Cisternas:2011}
{Cisternas}, M., {Jahnke}, K., {Inskip}, K.~J., {et~al.} 2011, \apj, 726, 57,
  \dodoi{10.1088/0004-637X/726/2/57}

\bibitem[{{Comerford} {et~al.}(2017){Comerford}, {Barrows},
  {M{\"u}ller-S{\'a}nchez}, {Nevin}, {Greene}, {Pooley}, {Stern}, \&
  {Harrison}}]{Comerford:2017b}
{Comerford}, J.~M., {Barrows}, R.~S., {M{\"u}ller-S{\'a}nchez}, F., {et~al.}
  2017, \apj, 849, 102, \dodoi{10.3847/1538-4357/aa8e4b}

\bibitem[{{Comerford} {et~al.}(2009){Comerford}, {Griffith}, {Gerke}, {Cooper},
  {Newman}, {Davis}, \& {Stern}}]{Comerford2009b}
{Comerford}, J.~M., {Griffith}, R.~L., {Gerke}, B.~F., {et~al.} 2009, \apjl,
  702, L82, \dodoi{10.1088/0004-637X/702/1/L82}

\bibitem[{{Comerford} {et~al.}(2015){Comerford}, {Pooley}, {Barrows}, {Greene},
  {Zakamska}, {Madejski}, \& {Cooper}}]{Comerford:2015}
{Comerford}, J.~M., {Pooley}, D., {Barrows}, R.~S., {et~al.} 2015, \apj, 806,
  219, \dodoi{10.1088/0004-637X/806/2/219}

\bibitem[{{Comerford} {et~al.}(2024){Comerford}, {Nevin}, {Negus}, {Barrows},
  {Eracleous}, {M{\"u}ller-S{\'a}nchez}, {Roy}, {Stemo}, {Storchi-Bergmann}, \&
  {Wylezalek}}]{Comerford:2024}
{Comerford}, J.~M., {Nevin}, R., {Negus}, J., {et~al.} 2024, \apj, 963, 53,
  \dodoi{10.3847/1538-4357/ad1a15}

\bibitem[{{Cox} {et~al.}(2008){Cox}, {Jonsson}, {Somerville}, {Primack}, \&
  {Dekel}}]{Cox:2008}
{Cox}, T.~J., {Jonsson}, P., {Somerville}, R.~S., {Primack}, J.~R., \& {Dekel},
  A. 2008, \mnras, 384, 386, \dodoi{10.1111/j.1365-2966.2007.12730.x}

\bibitem[{{Decarli} {et~al.}(2010){Decarli}, {Dotti}, {Montuori}, {Liimets}, \&
  {Ederoclite}}]{Decarli2010}
{Decarli}, R., {Dotti}, M., {Montuori}, C., {Liimets}, T., \& {Ederoclite}, A.
  2010, \apjl, 720, L93, \dodoi{10.1088/2041-8205/720/1/L93}

\bibitem[{{Di Matteo} {et~al.}(2005){Di Matteo}, {Springel}, \&
  {Hernquist}}]{DiMatteo:2005}
{Di Matteo}, T., {Springel}, V., \& {Hernquist}, L. 2005, \nat, 433, 604,
  \dodoi{10.1038/nature03335}

\bibitem[{{Dotti} {et~al.}(2009){Dotti}, {Montuori}, {Decarli}, {Volonteri},
  {Colpi}, \& {Haardt}}]{Dotti09}
{Dotti}, M., {Montuori}, C., {Decarli}, R., {et~al.} 2009, MNRA, 398, L73,
  \dodoi{10.1111/j.1745-3933.2009.00714.x}

\bibitem[{{Duncan} {et~al.}(2019){Duncan}, {Conselice}, {Mundy}, {Bell},
  {Donley}, {Galametz}, {Guo}, {Grogin}, {Hathi}, {Kartaltepe}, {Kocevski},
  {Koekemoer}, {P{\'e}rez-Gonz{\'a}lez}, {Mantha}, {Snyder}, \&
  {Stefanon}}]{Duncan:2019}
{Duncan}, K., {Conselice}, C.~J., {Mundy}, C., {et~al.} 2019, \apj, 876, 110,
  \dodoi{10.3847/1538-4357/ab148a}

\bibitem[{{Ellison} {et~al.}(2008){Ellison}, {Patton}, {Simard}, \&
  {McConnachie}}]{Ellison:2008}
{Ellison}, S.~L., {Patton}, D.~R., {Simard}, L., \& {McConnachie}, A.~W. 2008,
  \aj, 135, 1877, \dodoi{10.1088/0004-6256/135/5/1877}

\bibitem[{{Eracleous} {et~al.}(2012){Eracleous}, {Boroson}, {Halpern}, \&
  {Liu}}]{Eracleous:2012}
{Eracleous}, M., {Boroson}, T.~A., {Halpern}, J.~P., \& {Liu}, J. 2012, \apjs,
  201, 23, \dodoi{10.1088/0067-0049/201/2/23}

\bibitem[{{Eracleous} \& {Halpern}(2003)}]{EH03}
{Eracleous}, M., \& {Halpern}, J.~P. 2003, ApJ, 599, 886,
  \dodoi{10.1086/379540}

\bibitem[{{Eracleous} {et~al.}(1995){Eracleous}, {Livio}, {Halpern}, \&
  {Storchi-Bergmann}}]{Eracleous95}
{Eracleous}, M., {Livio}, M., {Halpern}, J.~P., \& {Storchi-Bergmann}, T. 1995,
  ApJ, 438, 610, \dodoi{10.1086/175104}

\bibitem[{{Ferrarese} \& {Merritt}(2000)}]{Ferrarese2000}
{Ferrarese}, L., \& {Merritt}, D. 2000, \apjl, 539, L9, \dodoi{10.1086/312838}

\bibitem[{{Fu} {et~al.}(2018){Fu}, {Steffen}, {Gross}, {Dai}, {Isbell}, {Lin},
  {Wake}, {Xue}, {Bizyaev}, \& {Pan}}]{Fu:2018}
{Fu}, H., {Steffen}, J.~L., {Gross}, A.~C., {et~al.} 2018, \apj, 856, 93,
  \dodoi{10.3847/1538-4357/aab364}

\bibitem[{{Gao} {et~al.}(2020){Gao}, {Wang}, {Pearson}, {Gordon}, {Holwerda},
  {Hopkins}, {Brown}, {Bland-Hawthorn}, \& {Owers}}]{Gao:2020}
{Gao}, F., {Wang}, L., {Pearson}, W.~J., {et~al.} 2020, \aap, 637, A94,
  \dodoi{10.1051/0004-6361/201937178}

\bibitem[{{Gaskell}(2009)}]{Gaskell:2009}
{Gaskell}, C.~M. 2009, \nar, 53, 140, \dodoi{10.1016/j.newar.2009.09.006}

\bibitem[{{Gebhardt} {et~al.}(2000){Gebhardt}, {Bender}, {Bower}, {Dressler},
  {Faber}, {Filippenko}, {Green}, {Grillmair}, {Ho}, {Kormendy}, {Lauer},
  {Magorrian}, {Pinkney}, {Richstone}, \& {Tremaine}}]{Gebhardt00}
{Gebhardt}, K., {Bender}, R., {Bower}, G., {et~al.} 2000, ApJ, 539, L13,
  \dodoi{10.1086/312840}

\bibitem[{{Goulding} {et~al.}(2018){Goulding}, {Greene}, {Bezanson}, {Greco},
  {Johnson}, {Leauthaud}, {Matsuoka}, {Medezinski}, \&
  {Price-Whelan}}]{Goulding:2018}
{Goulding}, A.~D., {Greene}, J.~E., {Bezanson}, R., {et~al.} 2018, \pasj, 70,
  S37, \dodoi{10.1093/pasj/psx135}

\bibitem[{{Gualandris} \& {Merritt}(2008)}]{Gualandris:2008}
{Gualandris}, A., \& {Merritt}, D. 2008, \apj, 678, 780, \dodoi{10.1086/586877}

\bibitem[{{G{\"u}ltekin} {et~al.}(2009){G{\"u}ltekin}, {Richstone}, {Gebhardt},
  {Lauer}, {Tremaine}, {Aller}, {Bender}, {Dressler}, {Faber}, {Filippenko},
  {Green}, {Ho}, {Kormendy}, {Magorrian}, {Pinkney}, \&
  {Siopis}}]{Gultekin:2009}
{G{\"u}ltekin}, K., {Richstone}, D.~O., {Gebhardt}, K., {et~al.} 2009, \apj,
  698, 198, \dodoi{10.1088/0004-637X/698/1/198}

\bibitem[{{Hao} {et~al.}(2005){Hao}, {Strauss}, {Tremonti}, {Schlegel},
  {Heckman}, {Kauffmann}, {Blanton}, {Fan}, {Gunn}, {Hall}, {Ivezi{\'c}},
  {Knapp}, {Krolik}, {Lupton}, {Richards}, {Schneider}, {Strateva}, {Zakamska},
  {Brinkmann}, {Brunner}, \& {Szokoly}}]{Hao:2005}
{Hao}, L., {Strauss}, M.~A., {Tremonti}, C.~A., {et~al.} 2005, \aj, 129, 1783,
  \dodoi{10.1086/428485}

\bibitem[{{Hickox} {et~al.}(2009){Hickox}, {Jones}, {Forman}, {Murray},
  {Kochanek}, {Eisenstein}, {Jannuzi}, {Dey}, {Brown}, {Stern}, {Eisenhardt},
  {Gorjian}, {Brodwin}, {Narayan}, {Cool}, {Kenter}, {Caldwell}, \&
  {Anderson}}]{Hickox:2009}
{Hickox}, R.~C., {Jones}, C., {Forman}, W.~R., {et~al.} 2009, \apj, 696, 891,
  \dodoi{10.1088/0004-637X/696/1/891}

\bibitem[{{Hinshaw} {et~al.}(2013){Hinshaw}, {Larson}, {Komatsu}, {Spergel},
  {Bennett}, {Dunkley}, {Nolta}, {Halpern}, {Hill}, {Odegard}, {Page}, {Smith},
  {Weiland}, {Gold}, {Jarosik}, {Kogut}, {Limon}, {Meyer}, {Tucker}, {Wollack},
  \& {Wright}}]{Hinshaw:2013}
{Hinshaw}, G., {Larson}, D., {Komatsu}, E., {et~al.} 2013, The Astrophysical
  Journal Supplement Series, 208, 19, \dodoi{10.1088/0067-0049/208/2/19}

\bibitem[{{Hoffman} \& {Loeb}(2007)}]{Hoffman:2007}
{Hoffman}, L., \& {Loeb}, A. 2007, \mnras, 377, 957,
  \dodoi{10.1111/j.1365-2966.2007.11694.x}

\bibitem[{{Hopkins} {et~al.}(2008){Hopkins}, {Hernquist}, {Cox}, \& {Kere{\v
  s}}}]{Hopkins2008}
{Hopkins}, P.~F., {Hernquist}, L., {Cox}, T.~J., \& {Kere{\v s}}, D. 2008,
  \apjs, 175, 356, \dodoi{10.1086/524362}

\bibitem[{{Hou} {et~al.}(2022){Hou}, {Li}, {Liu}, {Li}, {Li}, {Wang}, {Wang},
  \& {Ho}}]{Hou:2022}
{Hou}, M., {Li}, Z., {Liu}, X., {et~al.} 2022, arXiv e-prints,
  arXiv:2212.06399.
\newblock \doarXiv{2212.06399}

\bibitem[{{Jadhav} {et~al.}(2021){Jadhav}, {Robinson}, {Almeyda}, {Curran}, \&
  {Marconi}}]{Jadhav:2021}
{Jadhav}, Y., {Robinson}, A., {Almeyda}, T., {Curran}, R., \& {Marconi}, A.
  2021, \mnras, 507, 484, \dodoi{10.1093/mnras/stab2176}

\bibitem[{{Kalfountzou} {et~al.}(2017){Kalfountzou}, {Santos Lleo}, \&
  {Trichas}}]{Kalfountzou:2017}
{Kalfountzou}, E., {Santos Lleo}, M., \& {Trichas}, M. 2017, \apjl, 851, L15,
  \dodoi{10.3847/2041-8213/aa9b2d}

\bibitem[{{Keel} {et~al.}(2015){Keel}, {Maksym}, {Bennert}, {Lintott},
  {Chojnowski}, {Moiseev}, {Smirnova}, {Schawinski}, {Urry}, {Evans},
  {Pancoast}, {Scott}, {Showley}, \& {Flatland}}]{Keel:2015}
{Keel}, W.~C., {Maksym}, W.~P., {Bennert}, V.~N., {et~al.} 2015, \aj, 149, 155,
  \dodoi{10.1088/0004-6256/149/5/155}

\bibitem[{{Kewley} {et~al.}(2006){Kewley}, {Groves}, {Kauffmann}, \&
  {Heckman}}]{Kewley:2006}
{Kewley}, L.~J., {Groves}, B., {Kauffmann}, G., \& {Heckman}, T. 2006, \mnras,
  372, 961, \dodoi{10.1111/j.1365-2966.2006.10859.x}

\bibitem[{{Kocevski} {et~al.}(2012){Kocevski}, {Faber}, {Mozena}, {Koekemoer},
  {Nandra}, {Rangel}, {Laird}, {Brusa}, {Wuyts}, {Trump}, {Koo}, {Somerville},
  {Bell}, {Lotz}, {Alexander}, {Bournaud}, {Conselice}, {Dahlen}, {Dekel},
  {Donley}, {Dunlop}, {Finoguenov}, {Georgakakis}, {Giavalisco}, {Guo},
  {Grogin}, {Hathi}, {Juneau}, {Kartaltepe}, {Lucas}, {McGrath}, {McIntosh},
  {Mobasher}, {Robaina}, {Rosario}, {Straughn}, {van der Wel}, \&
  {Villforth}}]{Kocevski:2012}
{Kocevski}, D.~D., {Faber}, S.~M., {Mozena}, M., {et~al.} 2012, \apj, 744, 148,
  \dodoi{10.1088/0004-637X/744/2/148}

\bibitem[{{Kokubo} {et~al.}(2019){Kokubo}, {Mitsuda}, {Morokuma}, {Tominaga},
  {Tanaka}, {Moriya}, {Yoachim}, {Ivezi{\'c}}, {Sako}, \& {Doi}}]{Kokubo:2019}
{Kokubo}, M., {Mitsuda}, K., {Morokuma}, T., {et~al.} 2019, \apj, 872, 135,
  \dodoi{10.3847/1538-4357/aaff6b}

\bibitem[{{Komossa}(2012)}]{Komossa:2012}
{Komossa}, S. 2012, Advances in Astronomy, 2012, 364973,
  \dodoi{10.1155/2012/364973}

\bibitem[{{Komossa} {et~al.}(2003){Komossa}, {Burwitz}, {Hasinger}, {Predehl},
  {Kaastra}, \& {Ikebe}}]{Komossa2003}
{Komossa}, S., {Burwitz}, V., {Hasinger}, G., {et~al.} 2003, \apjl, 582, L15,
  \dodoi{10.1086/346145}

\bibitem[{{Komossa} \& {Merritt}(2008{\natexlab{a}})}]{Komossa:2008}
{Komossa}, S., \& {Merritt}, D. 2008{\natexlab{a}}, \apjl, 689, L89,
  \dodoi{10.1086/595883}

\bibitem[{{Komossa} \& {Merritt}(2008{\natexlab{b}})}]{Komossa:2008c}
---. 2008{\natexlab{b}}, \apjl, 683, L21, \dodoi{10.1086/591420}

\bibitem[{{Komossa} {et~al.}(2008{\natexlab{a}}){Komossa}, {Xu}, {Zhou},
  {Storchi-Bergmann}, \& {Binette}}]{Komossa:2008b}
{Komossa}, S., {Xu}, D., {Zhou}, H., {Storchi-Bergmann}, T., \& {Binette}, L.
  2008{\natexlab{a}}, \apj, 680, 926, \dodoi{10.1086/587932}

\bibitem[{{Komossa} {et~al.}(2008{\natexlab{b}}){Komossa}, {Zhou}, \&
  {Lu}}]{Komossa08a}
{Komossa}, S., {Zhou}, H., \& {Lu}, H. 2008{\natexlab{b}}, ApJ, 678, L81,
  \dodoi{10.1086/588656}

\bibitem[{{Koss} {et~al.}(2012){Koss}, {Mushotzky}, {Treister}, {Veilleux},
  {Vasudevan}, \& {Trippe}}]{Koss:2012}
{Koss}, M., {Mushotzky}, R., {Treister}, E., {et~al.} 2012, \apjl, 746, L22,
  \dodoi{10.1088/2041-8205/746/2/L22}

\bibitem[{{Koss} {et~al.}(2014){Koss}, {Blecha}, {Mushotzky}, {Hung},
  {Veilleux}, {Trakhtenbrot}, {Schawinski}, {Stern}, {Smith}, {Li}, {Man},
  {Filippenko}, {Mauerhan}, {Stanek}, \& {Sanders}}]{Koss:2014}
{Koss}, M., {Blecha}, L., {Mushotzky}, R., {et~al.} 2014, \mnras, 445, 515,
  \dodoi{10.1093/mnras/stu1673}

\bibitem[{{Kuncarayakti} {et~al.}(2023){Kuncarayakti}, {Sollerman}, {Izzo},
  {Maeda}, {Yang}, {Schulze}, {Angus}, {Aubert}, {Auchettl}, {Della Valle},
  {Dessart}, {Hinds}, {Kankare}, {Kawabata}, {Lundqvist}, {Nakaoka}, {Perley},
  {Raimundo}, {Strotjohann}, {Taguchi}, {Cai}, {Charalampopoulos}, {Fang},
  {Fraser}, {Guti{\'e}rrez}, {Imazawa}, {Kangas}, {Kawabata}, {Kotak},
  {Kravtsov}, {Matilainen}, {Mattila}, {Moran}, {Murata}, {Salmaso},
  {Anderson}, {Ashall}, {Bellm}, {Benetti}, {Chambers}, {Chen}, {Coughlin}, {De
  Colle}, {Fremling}, {Galbany}, {Gal-Yam}, {Gromadzki}, {Groom}, {Hajela},
  {Inserra}, {Kasliwal}, {Mahabal}, {Martin-Carrillo}, {Moore},
  {M{\"u}ller-Bravo}, {Nicholl}, {Ragosta}, {Riddle}, {Sharma}, {Srivastav},
  {Stritzinger}, {Wold}, \& {Young}}]{Kuncarayakti:2023}
{Kuncarayakti}, H., {Sollerman}, J., {Izzo}, L., {et~al.} 2023, \aap, 678,
  A209, \dodoi{10.1051/0004-6361/202346526}

\bibitem[{{Lacerda} {et~al.}(2022){Lacerda}, {S{\'a}nchez},
  {Mej{\'\i}a-Narv{\'a}ez}, {Camps-Fari{\~n}a}, {Espinosa-Ponce},
  {Barrera-Ballesteros}, {Ibarra-Medel}, \& {Lugo-Aranda}}]{Lacerda:2022}
{Lacerda}, E. A.~D., {S{\'a}nchez}, S.~F., {Mej{\'\i}a-Narv{\'a}ez}, A.,
  {et~al.} 2022, \na, 97, 101895, \dodoi{10.1016/j.newast.2022.101895}

\bibitem[{{LaMassa} {et~al.}(2015){LaMassa}, {Cales}, {Moran}, {Myers},
  {Richards}, {Eracleous}, {Heckman}, {Gallo}, \& {Urry}}]{LaMassa:2015}
{LaMassa}, S.~M., {Cales}, S., {Moran}, E.~C., {et~al.} 2015, \apj, 800, 144,
  \dodoi{10.1088/0004-637X/800/2/144}

\bibitem[{{Law} {et~al.}(2016){Law}, {Cherinka}, {Yan}, {Andrews}, {Bershady},
  {Bizyaev}, {Blanc}, {Blanton}, {Bolton}, {Brownstein}, {Bundy}, {Chen},
  {Drory}, {D'Souza}, {Fu}, {Jones}, {Kauffmann}, {MacDonald}, {Masters},
  {Newman}, {Parejko}, {S{\'a}nchez-Gallego}, {S{\'a}nchez}, {Schlegel},
  {Thomas}, {Wake}, {Weijmans}, {Westfall}, \& {Zhang}}]{Law:2016}
{Law}, D.~R., {Cherinka}, B., {Yan}, R., {et~al.} 2016, \aj, 152, 83,
  \dodoi{10.3847/0004-6256/152/4/83}

\bibitem[{{Lena} {et~al.}(2020){Lena}, {Jonker}, {Rauer}, {Hernandez}, \&
  {Kostrzewa-Rutkowska}}]{Lena:2020}
{Lena}, D., {Jonker}, P.~G., {Rauer}, J.~P., {Hernandez}, S., \&
  {Kostrzewa-Rutkowska}, Z. 2020, \mnras, 495, 1771,
  \dodoi{10.1093/mnras/staa1174}

\bibitem[{{Li} {et~al.}(2012){Li}, {Liu}, {Berczik}, {Chen}, \&
  {Spurzem}}]{Li:2012}
{Li}, S., {Liu}, F.~K., {Berczik}, P., {Chen}, X., \& {Spurzem}, R. 2012, \apj,
  748, 65, \dodoi{10.1088/0004-637X/748/1/65}

\bibitem[{{Liu} {et~al.}(2019){Liu}, {Liu}, {Dong}, {Zhou}, {Wang}, {Lu}, \&
  {Yuan}}]{Liu:2019}
{Liu}, H.-Y., {Liu}, W.-J., {Dong}, X.-B., {et~al.} 2019, \apjs, 243, 21,
  \dodoi{10.3847/1538-4365/ab298b}

\bibitem[{{Lotz} {et~al.}(2011){Lotz}, {Jonsson}, {Cox}, {Croton}, {Primack},
  {Somerville}, \& {Stewart}}]{Lotz:2011}
{Lotz}, J.~M., {Jonsson}, P., {Cox}, T.~J., {et~al.} 2011, \apj, 742, 103,
  \dodoi{10.1088/0004-637X/742/2/103}

\bibitem[{{Lousto} \& {Zlochower}(2011)}]{Lousto:2011}
{Lousto}, C.~O., \& {Zlochower}, Y. 2011, \prd, 83, 024003,
  \dodoi{10.1103/PhysRevD.83.024003}

\bibitem[{{Lyke} {et~al.}(2020){Lyke}, {Higley}, {McLane}, {Schurhammer},
  {Myers}, {Ross}, {Dawson}, {Chabanier}, {Martini}, {Busca}, {Mas des
  Bourboux}, {Salvato}, {Streblyanska}, {Zarrouk}, {Burtin}, {Anderson},
  {Bautista}, {Bizyaev}, {Brandt}, {Brinkmann}, {Brownstein}, {Comparat},
  {Green}, {de la Macorra}, {Mu{\~n}oz Guti{\'e}rrez}, {Hou}, {Newman},
  {Palanque-Delabrouille}, {P{\^a}ris}, {Percival}, {Petitjean}, {Rich},
  {Rossi}, {Schneider}, {Smith}, {Vivek}, \& {Weaver}}]{Lyke:2020}
{Lyke}, B.~W., {Higley}, A.~N., {McLane}, J.~N., {et~al.} 2020, \apjs, 250, 8,
  \dodoi{10.3847/1538-4365/aba623}

\bibitem[{{Markakis} {et~al.}(2015){Markakis}, {Dierkes}, {Eckart},
  {Nishiyama}, {Britzen}, {Garc{\'\i}a-Mar{\'\i}n}, {Horrobin}, {Muxlow}, \&
  {Zensus}}]{Markakis:2015}
{Markakis}, K., {Dierkes}, J., {Eckart}, A., {et~al.} 2015, \aap, 580, A11,
  \dodoi{10.1051/0004-6361/201425077}

\bibitem[{{Mazzarella} {et~al.}(2012){Mazzarella}, {Iwasawa}, {Vavilkin},
  {Armus}, {Kim}, {Bothun}, {Evans}, {Spoon}, {Haan}, {Howell}, {Lord},
  {Marshall}, {Ishida}, {Xu}, {Petric}, {Sanders}, {Surace}, {Appleton},
  {Chan}, {Frayer}, {Inami}, {Khachikian}, {Madore}, {Privon}, {Sturm}, {U}, \&
  {Veilleux}}]{Mazzarella:2012}
{Mazzarella}, J.~M., {Iwasawa}, K., {Vavilkin}, T., {et~al.} 2012, \aj, 144,
  125, \dodoi{10.1088/0004-6256/144/5/125}

\bibitem[{{McAlpine} {et~al.}(2020){McAlpine}, {Harrison}, {Rosario},
  {Alexander}, {Ellison}, {Johansson}, \& {Patton}}]{McAlpine:2020}
{McAlpine}, S., {Harrison}, C.~M., {Rosario}, D.~J., {et~al.} 2020, \mnras,
  494, 5713, \dodoi{10.1093/mnras/staa1123}

\bibitem[{{Meena} {et~al.}(2021){Meena}, {Crenshaw}, {Schmitt}, {Revalski},
  {Fischer}, {Polack}, {Kraemer}, \& {Dashtamirova}}]{Meena:2021}
{Meena}, B., {Crenshaw}, D.~M., {Schmitt}, H.~R., {et~al.} 2021, \apj, 916, 31,
  \dodoi{10.3847/1538-4357/ac0246}

\bibitem[{{Merritt} {et~al.}(2009){Merritt}, {Schnittman}, \&
  {Komossa}}]{Merritt:2009}
{Merritt}, D., {Schnittman}, J.~D., \& {Komossa}, S. 2009, \apj, 699, 1690,
  \dodoi{10.1088/0004-637X/699/2/1690}

\bibitem[{{M{\"u}ller-S{\'a}nchez} {et~al.}(2015){M{\"u}ller-S{\'a}nchez},
  {Comerford}, {Nevin}, {Barrows}, {Cooper}, \&
  {Greene}}]{Mueller-Sanchez:2015}
{M{\"u}ller-S{\'a}nchez}, F., {Comerford}, J.~M., {Nevin}, R., {et~al.} 2015,
  \apj, 813, 103, \dodoi{10.1088/0004-637X/813/2/103}

\bibitem[{{Negus} {et~al.}(2024){Negus}, {Comerford}, \& {M{\"u}ller
  S{\'a}nchez}}]{Negus:2024}
{Negus}, J., {Comerford}, J.~M., \& {M{\"u}ller S{\'a}nchez}, F. 2024, \apj,
  971, 92, \dodoi{10.3847/1538-4357/ad4c68}

\bibitem[{{Nevin} {et~al.}(2023){Nevin}, {Blecha}, {Comerford}, {Simon},
  {Terrazas}, {Barrows}, \& {V{\'a}zquez-Mata}}]{Nevin:2023}
{Nevin}, R., {Blecha}, L., {Comerford}, J., {et~al.} 2023, \mnras, 522, 1,
  \dodoi{10.1093/mnras/stad911}

\bibitem[{{Oh} {et~al.}(2015){Oh}, {Yi}, {Schawinski}, {Koss}, {Trakhtenbrot},
  \& {Soto}}]{Oh:2015}
{Oh}, K., {Yi}, S.~K., {Schawinski}, K., {et~al.} 2015, \apjs, 219, 1,
  \dodoi{10.1088/0067-0049/219/1/1}

\bibitem[{{Peng} {et~al.}(2010){Peng}, {Ho}, {Impey}, \& {Rix}}]{Peng:2010}
{Peng}, C.~Y., {Ho}, L.~C., {Impey}, C.~D., \& {Rix}, H.-W. 2010, \aj, 139,
  2097, \dodoi{10.1088/0004-6256/139/6/2097}

\bibitem[{{Peres}(1962)}]{Peres:1962}
{Peres}, A. 1962, Physical Review, 128, 2471, \dodoi{10.1103/PhysRev.128.2471}

\bibitem[{{Piconcelli} {et~al.}(2010){Piconcelli}, {Vignali}, {Bianchi},
  {Mathur}, {Fiore}, {Guainazzi}, {Lanzuisi}, {Maiolino}, \&
  {Nicastro}}]{Piconcelli2010}
{Piconcelli}, E., {Vignali}, C., {Bianchi}, S., {et~al.} 2010, \apjl, 722,
  L147, \dodoi{10.1088/2041-8205/722/2/L147}

\bibitem[{{Robinson} {et~al.}(2010){Robinson}, {Young}, {Axon}, {Kharb}, \&
  {Smith}}]{Robinson:2010}
{Robinson}, A., {Young}, S., {Axon}, D.~J., {Kharb}, P., \& {Smith}, J.~E.
  2010, \apjl, 717, L122, \dodoi{10.1088/2041-8205/717/2/L122}

\bibitem[{{Runnoe} {et~al.}(2015){Runnoe}, {Eracleous}, {Mathes}, {Pennell},
  {Boroson}, {Sigur{\dh}sson}, {Bogdanovi{\'c}}, {Halpern}, \&
  {Liu}}]{Runnoe:2015}
{Runnoe}, J.~C., {Eracleous}, M., {Mathes}, G., {et~al.} 2015, \apjs, 221, 7,
  \dodoi{10.1088/0067-0049/221/1/7}

\bibitem[{{Runnoe} {et~al.}(2016){Runnoe}, {Cales}, {Ruan}, {Eracleous},
  {Anderson}, {Shen}, {Green}, {Morganson}, {LaMassa}, {Greene}, {Dwelly},
  {Schneider}, {Merloni}, {Georgakakis}, \& {Roman-Lopes}}]{Runnoe:2016}
{Runnoe}, J.~C., {Cales}, S., {Ruan}, J.~J., {et~al.} 2016, \mnras, 455, 1691,
  \dodoi{10.1093/mnras/stv2385}

\bibitem[{{Runnoe} {et~al.}(2017){Runnoe}, {Eracleous}, {Pennell}, {Mathes},
  {Boroson}, {Sigurdsson}, {Bogdanovic}, {Halpern}, {Liu}, \&
  {Brown}}]{Runnoe:2017}
{Runnoe}, J.~C., {Eracleous}, M., {Pennell}, A., {et~al.} 2017, ArXiv e-prints.
\newblock \doarXiv{1702.05465}

\bibitem[{{S{\'a}nchez} {et~al.}(2022){S{\'a}nchez}, {Barrera-Ballesteros},
  {Lacerda}, {Mej{\'\i}a-Narvaez}, {Camps-Fari{\~n}a}, {Bruzual},
  {Espinosa-Ponce}, {Rodr{\'\i}guez-Puebla}, {Calette}, {Ibarra-Medel},
  {Avila-Reese}, {Hernandez-Toledo}, {Bershady}, {Cano-Diaz}, \&
  {Munguia-Cordova}}]{Sanchez:2022}
{S{\'a}nchez}, S.~F., {Barrera-Ballesteros}, J.~K., {Lacerda}, E., {et~al.}
  2022, \apjs, 262, 36, \dodoi{10.3847/1538-4365/ac7b8f}

\bibitem[{{Satyapal} {et~al.}(2014){Satyapal}, {Ellison}, {McAlpine}, {Hickox},
  {Patton}, \& {Mendel}}]{Satyapal:2014}
{Satyapal}, S., {Ellison}, S.~L., {McAlpine}, W., {et~al.} 2014, \mnras, 441,
  1297, \dodoi{10.1093/mnras/stu650}

\bibitem[{{Schawinski} {et~al.}(2015){Schawinski}, {Koss}, {Berney}, \&
  {Sartori}}]{Schawinski:2015}
{Schawinski}, K., {Koss}, M., {Berney}, S., \& {Sartori}, L.~F. 2015, \mnras,
  451, 2517, \dodoi{10.1093/mnras/stv1136}

\bibitem[{{Schneider} {et~al.}(2010){Schneider}, {Richards}, {Hall}, {Strauss},
  {Anderson}, {Boroson}, {Ross}, {Shen}, {Brandt}, {Fan}, {Inada}, {Jester},
  {Knapp}, {Krawczyk}, {Thakar}, {Vanden Berk}, {Voges}, {Yanny}, {York},
  {Bahcall}, {Bizyaev}, {Blanton}, {Brewington}, {Brinkmann}, {Eisenstein},
  {Frieman}, {Fukugita}, {Gray}, {Gunn}, {Hibon}, {Ivezi{\'c}}, {Kent}, {Kron},
  {Lee}, {Lupton}, {Malanushenko}, {Malanushenko}, {Oravetz}, {Pan}, {Pier},
  {Price}, {Saxe}, {Schlegel}, {Simmons}, {Snedden}, {SubbaRao}, {Szalay}, \&
  {Weinberg}}]{Schneider:2010}
{Schneider}, D.~P., {Richards}, G.~T., {Hall}, P.~B., {et~al.} 2010, \aj, 139,
  2360, \dodoi{10.1088/0004-6256/139/6/2360}

\bibitem[{{Schnittman} \& {Buonanno}(2007)}]{Schnittman:2007}
{Schnittman}, J.~D., \& {Buonanno}, A. 2007, \apjl, 662, L63,
  \dodoi{10.1086/519309}

\bibitem[{{Shen} {et~al.}(2011){Shen}, {Richards}, {Strauss}, {Hall},
  {Schneider}, {Snedden}, {Bizyaev}, {Brewington}, {Malanushenko},
  {Malanushenko}, {Oravetz}, {Pan}, \& {Simmons}}]{Shen:2011a}
{Shen}, Y., {Richards}, G.~T., {Strauss}, M.~A., {et~al.} 2011, \apjs, 194, 45,
  \dodoi{10.1088/0067-0049/194/2/45}

\bibitem[{{Shields} {et~al.}(2009){Shields}, {Rosario}, {Smith}, {Bonning},
  {Salviander}, {Kalirai}, {Strickler}, {Ramirez-Ruiz}, {Dutton}, {Treu}, \&
  {Marshall}}]{Shields09}
{Shields}, G.~A., {Rosario}, D.~J., {Smith}, K.~L., {et~al.} 2009, ApJ, 707,
  936, \dodoi{10.1088/0004-637X/707/2/936}

\bibitem[{{Sijacki} {et~al.}(2011){Sijacki}, {Springel}, \&
  {Haehnelt}}]{Sijacki:2011}
{Sijacki}, D., {Springel}, V., \& {Haehnelt}, M.~G. 2011, \mnras, 414, 3656,
  \dodoi{10.1111/j.1365-2966.2011.18666.x}

\bibitem[{{Skipper} \& {Browne}(2018)}]{Skipper:2018}
{Skipper}, C.~J., \& {Browne}, I.~W.~A. 2018, \mnras, 475, 5179,
  \dodoi{10.1093/mnras/sty114}

\bibitem[{{Sofue} \& {Rubin}(2001)}]{Sofue:2001}
{Sofue}, Y., \& {Rubin}, V. 2001, \araa, 39, 137,
  \dodoi{10.1146/annurev.astro.39.1.137}

\bibitem[{{Somerville} {et~al.}(2001){Somerville}, {Primack}, \&
  {Faber}}]{Somerville:2001}
{Somerville}, R.~S., {Primack}, J.~R., \& {Faber}, S.~M. 2001, \mnras, 320,
  504, \dodoi{10.1046/j.1365-8711.2001.03975.x}

\bibitem[{{Springel} {et~al.}(2005){Springel}, {Di Matteo}, \&
  {Hernquist}}]{Springel:2005a}
{Springel}, V., {Di Matteo}, T., \& {Hernquist}, L. 2005, \apjl, 620, L79,
  \dodoi{10.1086/428772}

\bibitem[{{Steffen} {et~al.}(2022){Steffen}, {Fu}, {Brownstein}, {Comerford},
  {Cruz-Gonz{\'a}lez}, {Dai}, {Drory}, {Gross}, {Negrete}, \&
  {Yan}}]{Steffen:2022}
{Steffen}, J.~L., {Fu}, H., {Brownstein}, J.~R., {et~al.} 2022, arXiv e-prints,
  arXiv:2212.02677.
\newblock \doarXiv{2212.02677}

\bibitem[{{Steinborn} {et~al.}(2016){Steinborn}, {Dolag}, {Comerford},
  {Hirschmann}, {Remus}, \& {Teklu}}]{Steinborn:2016}
{Steinborn}, L.~K., {Dolag}, K., {Comerford}, J.~M., {et~al.} 2016, \mnras,
  458, 1013, \dodoi{10.1093/mnras/stw316}

\bibitem[{{Steinhardt} {et~al.}(2012){Steinhardt}, {Schramm}, {Silverman},
  {Alexandroff}, {Capak}, {Civano}, {Elvis}, {Masters}, {Mobasher},
  {Pattarakijwanich}, \& {Strauss}}]{Steinhardt:2012}
{Steinhardt}, C.~L., {Schramm}, M., {Silverman}, J.~D., {et~al.} 2012, \apj,
  759, 24, \dodoi{10.1088/0004-637X/759/1/24}

\bibitem[{{Stemo} {et~al.}(2021){Stemo}, {Comerford}, {Barrows}, {Stern},
  {Assef}, {Griffith}, \& {Schechter}}]{Stemo:2021}
{Stemo}, A., {Comerford}, J.~M., {Barrows}, R.~S., {et~al.} 2021, \apj, 923,
  36, \dodoi{10.3847/1538-4357/ac0bbf}

\bibitem[{{Stern} \& {Laor}(2012)}]{Stern:Laor:2012}
{Stern}, J., \& {Laor}, A. 2012, \mnras, 423, 600,
  \dodoi{10.1111/j.1365-2966.2012.20901.x}

\bibitem[{{Sulentic} {et~al.}(2000){Sulentic}, {Marziani}, \&
  {Dultzin-Hacyan}}]{Sulentic:2000}
{Sulentic}, J.~W., {Marziani}, P., \& {Dultzin-Hacyan}, D. 2000, \araa, 38,
  521, \dodoi{10.1146/annurev.astro.38.1.521}

\bibitem[{{Tsai} {et~al.}(2013){Tsai}, {Jarrett}, {Stern}, {Emonts}, {Barrows},
  {Assef}, {Norris}, {Eisenhardt}, {Lonsdale}, {Blain}, {Benford}, {Wu},
  {Stalder}, {Stubbs}, {High}, {Li}, \& {Kong}}]{Tsai:2013}
{Tsai}, C.-W., {Jarrett}, T.~H., {Stern}, D., {et~al.} 2013, \apj, 779, 41,
  \dodoi{10.1088/0004-637X/779/1/41}

\bibitem[{{Valenti} {et~al.}(2018){Valenti}, {Zoccali}, {Mucciarelli},
  {Gonzalez}, {Surot}, {Minniti}, {Rejkuba}, {Pasquini}, {Fiorentino}, {Bono},
  {Rich}, \& {Soto}}]{Valenti:2018}
{Valenti}, E., {Zoccali}, M., {Mucciarelli}, A., {et~al.} 2018, \aap, 616, A83,
  \dodoi{10.1051/0004-6361/201832905}

\bibitem[{{Villforth} {et~al.}(2014){Villforth}, {Hamann}, {Rosario},
  {Santini}, {McGrath}, {van der Wel}, {Chang}, {Guo}, {Dahlen}, {Bell},
  {Conselice}, {Croton}, {Dekel}, {Faber}, {Grogin}, {Hamilton}, {Hopkins},
  {Juneau}, {Kartaltepe}, {Kocevski}, {Koekemoer}, {Koo}, {Lotz}, {McIntosh},
  {Mozena}, {Somerville}, \& {Wild}}]{Villforth:2014}
{Villforth}, C., {Hamann}, F., {Rosario}, D.~J., {et~al.} 2014, \mnras, 439,
  3342, \dodoi{10.1093/mnras/stu173}

\bibitem[{{Volonteri}(2007)}]{Volonteri:2007}
{Volonteri}, M. 2007, \apjl, 663, L5, \dodoi{10.1086/519525}

\bibitem[{Volonteri \& Madau(2008)}]{Volonteri:2008b}
Volonteri, M., \& Madau, P. 2008, The Astrophysical Journal, 687, L57

\bibitem[{{Weigel} {et~al.}(2018){Weigel}, {Schawinski}, {Treister},
  {Trakhtenbrot}, \& {Sanders}}]{Weigel:2018}
{Weigel}, A.~K., {Schawinski}, K., {Treister}, E., {Trakhtenbrot}, B., \&
  {Sanders}, D.~B. 2018, \mnras, 476, 2308, \dodoi{10.1093/mnras/sty383}

\bibitem[{{Weston} {et~al.}(2017){Weston}, {McIntosh}, {Brodwin}, {Mann},
  {Cooper}, {McConnell}, \& {Nielsen}}]{Weston:2017}
{Weston}, M.~E., {McIntosh}, D.~H., {Brodwin}, M., {et~al.} 2017, \mnras, 464,
  3882, \dodoi{10.1093/mnras/stw2620}

\bibitem[{{Wylezalek} {et~al.}(2020){Wylezalek}, {Flores}, {Zakamska},
  {Greene}, \& {Riffel}}]{Wylezalek:2020}
{Wylezalek}, D., {Flores}, A.~M., {Zakamska}, N.~L., {Greene}, J.~E., \&
  {Riffel}, R.~A. 2020, \mnras, 492, 4680, \dodoi{10.1093/mnras/staa062}

\bibitem[{{Wylezalek} {et~al.}(2018){Wylezalek}, {Zakamska}, {Greene},
  {Riffel}, {Drory}, {Andrews}, {Merloni}, \& {Thomas}}]{Wylezalek:2018}
{Wylezalek}, D., {Zakamska}, N.~L., {Greene}, J.~E., {et~al.} 2018, \mnras,
  474, 1499, \dodoi{10.1093/mnras/stx2784}

\bibitem[{{Yan} {et~al.}(2015){Yan}, {Quimby}, {Ofek}, {Gal-Yam}, {Mazzali},
  {Perley}, {Vreeswijk}, {Leloudas}, {De Cia}, {Masci}, {Cenko}, {Cao},
  {Kulkarni}, {Nugent}, {Rebbapragada}, {Wo{\'z}niak}, \& {Yaron}}]{Yan:2015}
{Yan}, L., {Quimby}, R., {Ofek}, E., {et~al.} 2015, \apj, 814, 108,
  \dodoi{10.1088/0004-637X/814/2/108}

\bibitem[{{Yu} {et~al.}(2011){Yu}, {Lu}, {Mohayaee}, \& {Colin}}]{Yu:2011}
{Yu}, Q., {Lu}, Y., {Mohayaee}, R., \& {Colin}, J. 2011, \apj, 738, 92,
  \dodoi{10.1088/0004-637X/738/1/92}

\bibitem[{{Yuan} {et~al.}(2018){Yuan}, {Argudo-Fern{\'a}ndez}, {Shen}, {Hao},
  {Jiang}, {Yin}, {Boquien}, \& {Lin}}]{Yuan:2018}
{Yuan}, F.-T., {Argudo-Fern{\'a}ndez}, M., {Shen}, S., {et~al.} 2018, \aap,
  613, A13, \dodoi{10.1051/0004-6361/201731865}

\end{thebibliography}
\end{document}